\documentclass[11pt,sort&compress]{elsarticle}

\makeatletter
  \long\def\pprintMaketitle{\clearpage
  \iflongmktitle\if@twocolumn\let\columnwidth=\textwidth\fi\fi
  \resetTitleCounters
  \def\baselinestretch{1}%
  \printFirstPageNotes
  \begin{center}%
 \thispagestyle{pprintTitle}%
   \def\baselinestretch{1}%
    {\large\bf\@title}\par\vskip5pt
    \normalsize\elsauthors\par\vskip5pt
    \footnotesize\itshape\elsaddress\par\vskip10pt
    \end{center}%
  \gdef\thefootnote{\arabic{footnote}}%
  }
\makeatother

\newcommand\blfootnote[1]{%
  \begingroup
  \renewcommand\thefootnote{}\footnote{#1}%
  \addtocounter{footnote}{-1}%
  \endgroup
}

\usepackage{standalone}
\usepackage[margin=1in]{geometry}
\usepackage[dvipsnames]{xcolor} 

\usepackage[]{siunitx}
\usepackage{array} 
\usepackage{graphicx} 
\usepackage{enumitem}
\usepackage[version=4]{mhchem}
\usepackage{setspace} 
\usepackage{subfig} 
\usepackage{stmaryrd}

\usepackage{placeins}
\usepackage{ulem}

\usepackage{hyperref}
\hypersetup{
  colorlinks=true,
}

\usepackage[font=small,justification=justified]{caption} 
\usepackage{float} 
\usepackage{newfloat}
\usepackage{subcaption}

\usepackage{xfrac}
\usepackage{amsmath} 
\usepackage{commath} 

\usepackage{amssymb} 
\usepackage{mathtools} 
\usepackage{bm} 

\usepackage{amsmath,amsfonts,amssymb,amsthm,mathrsfs,mathtools}

\definecolor{lightblue}{rgb}{0.63, 0.74, 0.78}
\definecolor{seagreen}{rgb}{0.18, 0.42, 0.41}
\definecolor{orange}{rgb}{0.85, 0.55, 0.13}
\definecolor{silver}{rgb}{0.69, 0.67, 0.66}
\definecolor{rust}{rgb}{0.72, 0.26, 0.06}
\definecolor{purp}{RGB}{68, 14, 156}

\colorlet{lightrust}{rust!50!white}
\colorlet{lightorange}{orange!25!white}
\colorlet{lightlightblue}{lightblue}
\colorlet{lightsilver}{silver!30!white}
\colorlet{darkorange}{orange!75!black}
\colorlet{darksilver}{silver!65!black}
\colorlet{darklightblue}{lightblue!65!black}
\colorlet{darkrust}{rust!85!black}
\colorlet{darkseagreen}{seagreen!85!black}

\usepackage[labelfont=bf,font=small]{caption}

\usepackage[parfill]{parskip}

\makeatletter 
\g@addto@macro\normalsize{%
  \setlength\abovedisplayskip{5pt}
  \setlength\belowdisplayskip{10pt}
  \setlength\abovedisplayshortskip{5pt}
  \setlength\belowdisplayshortskip{10pt}
}
\makeatother

\newcommand\doubleplus{+\kern-1.3ex+\kern0.8ex}

\newcommand{\pop}[2]{\frac{\partial#1}{\partial#2}}

\newcommand{\pp}[1]{\ensuremath{\left( #1 \right)}}
\newcommand{\psq}[1]{\ensuremath{{\left[ #1 \right]}}}

\usepackage{accents}
\newlength{\dhatheight}

\definecolor{RYB0}{RGB}{0.5,0.5,0.5}
\definecolor{RYB1}{RGB}{207, 37, 37}
\definecolor{RYB2}{RGB}{37, 91, 207}
\definecolor{RYB3}{RGB}{37, 207, 91}
\definecolor{RYB4}{RGB}{163,26,145}
\definecolor{RYB5}{RGB}{253, 180, 98}
\definecolor{RYB6}{RGB}{179, 222, 105}
\definecolor{RYB7}{RGB}{128, 177, 211}

\usepackage[colorinlistoftodos, color=blue!20!white, bordercolor=gray,
textsize=tiny,textwidth=0.8in,prependcaption]{todonotes}

\usepackage{listings}
\usepackage{parcolumns}

\lstdefinestyle{inline}{
language=Python,
basicstyle={\ttfamily\footnotesize},
keywordstyle=\color{RYB4},
otherkeywords={self, as},
morekeywords={self, as},
stringstyle=\color{RYB3!50!black},
showstringspaces=false,
tabsize=8,
captionpos=r,
escapechar={§},
}

\lstdefinestyle{float}{
language=Python,
basicstyle={\ttfamily\footnotesize},
keywordstyle=\color{RYB4},
otherkeywords={self, as},
morekeywords={self, as},
stringstyle=\color{RYB3!50!black},
frame=tb,
showstringspaces=false,
numbers=left,
numbersep=5pt,
captionpos=b,
tabsize=8,
showtabs=true,
}

\usepackage{tikz}
\usepackage{scalerel}

\definecolor{RYB1}{RGB}{207, 37, 37}
\definecolor{RYB2}{RGB}{37, 91, 207}
\definecolor{RYB3}{RGB}{37, 207, 91}
\definecolor{RYB4}{RGB}{163,26,145}
\definecolor{RYB5}{RGB}{253, 180, 98}
\definecolor{RYB6}{RGB}{179, 222, 105}
\definecolor{RYB7}{RGB}{128, 177, 211}
\definecolor{myBlue}{RGB}{19,41,75}  
\definecolor{myOrange}{RGB}{232,74,39}   
\definecolor{myBrown}{RGB}{155,77,40}
\definecolor{myTan}{RGB}{205,133,63}
\definecolor{mediumseagreen}{RGB}{60, 179, 113}

\def\Cpp{{C\nolinebreak[4]\hspace{-.05em}\raisebox{.4ex}{\tiny\bf ++}}}

\usepackage[nameinlink]{cleveref}
\crefname{equation}{}{}
\crefname{appendix}{}{}

\usepackage{outlines}
\usepackage{abstract}

\lstdefinestyle{custompython}{
    language=Python,
    backgroundcolor={\color{white}},
    basicstyle={\ttfamily\color{myBlue}},
    breakatwhitespace=true,
    breaklines=true,
    breakindent=7.5em,
    postbreak=\mbox{\textcolor{red}{$\hookrightarrow$}\space},
    captionpos=b,
    deletekeywords={...},
    escapeinside={"*}{*")},
    extendedchars=true,
    firstnumber=1,
    frame=tb,
    keepspaces=true,
    keywordstyle={\color{RYB4}},
    morekeywords={assert},
    numbers=left,
    numbersep=5pt,
    numberstyle={\tiny\color{myTan}},
    rulecolor={\color{black}},
    showspaces=false,
    showstringspaces=false,
    showtabs=false,
    stepnumber=1,
    stringstyle={\color{mediumseagreen}},
    tabsize=2,
    commentstyle=\color{myOrange},
    morecomment=[s]{"""}{"""},
    moredelim=[s][\color{teal}]{"""}{"""} 
}

\usepackage{float}
\restylefloat{table}
\DeclareFloatingEnvironment[fileext=frm,placement={!ht},name=Listing,within=section]{listing}

\begin{document}

\hypersetup{
  linkcolor=darkrust,
  citecolor=seagreen,
  urlcolor=darkrust,
  pdfauthor=author,
}

\begin{frontmatter}

\title{{\large\bf Pyrometheus: Symbolic abstractions for XPU and automatically differentiated computation of combustion kinetics and thermodynamics}}

\author[uiucMechse]{Esteban Cisneros-Garibay\corref{cor1}}
\ead{ecisnerosg88@gmail.com}
\cortext[cor1]{Corresponding author}
\author[gatech]{Henry~Le~Berre}
\author[gatech]{Dimitrios~Adam}
\author[gatech,gatechae,gatechme]{Spencer~H.~Bryngelson}
\author[uiucAero]{Jonathan~B.~Freund}

\address[uiucMechse]{Mechanical Science \& Engineering, University of Illinois Urbana--Champaign, Urbana, IL, USA\vspace{-0.15cm}}
\address[gatech]{School of Computational Science and Engineering, Georgia Institute of Technology, Atlanta, GA, USA\vspace{-0.15cm}} 
\address[gatechae]{Daniel Guggenheim School of Aerospace Engineering, Georgia Institute of Technology, Atlanta, GA, USA\vspace{-0.15cm}} 
\address[gatechme]{George W.\ Woodruff School of Mechanical Engineering, Georgia Institute of Technology, Atlanta, GA, USA\vspace{-0.15cm}} 
\address[uiucAero]{Aerospace Engineering, University of Illinois Urbana--Champaign, Urbana, IL, USA}

\date{}

\end{frontmatter}

\blfootnote{
\noindent Code available at: \url{https://github.com/pyrometheus} and \url{https://github.com/MFlowCode}
}

\begin{abstract}
The cost of combustion simulations is often dominated by the evaluation of net production rates of chemical species and mixture thermodynamics (thermochemistry).
Execution on computing accelerators (XPUs) like graphic processing units (GPUs) can greatly reduce this cost.
However, established thermochemistry software is not readily portable to such devices or sacrifices valuable analytical forms that enable differentiation for sensitivity analysis and implicit time integration.
Symbolic abstractions are developed with corresponding transformations that enable computation on accelerators and automatic differentiation by avoiding premature specification of detail.
The software package Pyrometheus is introduced as an implementation of these abstractions and their transformations for combustion thermochemistry.
The formulation facilitates code generation from the symbolic representation of a specific thermochemical mechanism in multiple target languages, including Python, \Cpp{}, and Fortran.
Computational concerns are separated: the generated code processes array-valued expressions but does not specify their semantics.
These semantics are provided by compatible array libraries, such as NumPy, Pytato, and Google JAX.
Thus, the generated code retains a symbolic representation of the thermochemistry, which translates to computation on accelerators and CPUs and automatic differentiation.
The design and operation of these symbolic abstractions and their companion tool, Pyrometheus, are discussed throughout.
Roofline demonstrations show that the computation of chemical source terms within MFC, a Fortran-based flow solver we link to Pyrometheus, is performant.
\end{abstract}

\section{Motivation and Significance}\label{sec:introduction}

\subsection{Computational Challenges of Combustion Thermochemistry}

Combustion underpins modern transportation and energy production, so tools to investigate and predict are useful.
Simulations of reacting flows promise to improve combustion, particularly when experiments are costly or challenging~\cite{ref:Alexander2020}.
The governing equations for reacting flows involve nonlinear chemical source terms and equations of state, which must be evaluated in such simulations.
Depending on the goals of the application, implementations need to satisfy different computational and analytical properties~\cite{ref:Henry2022}.
This work introduces computational strategies and tools that enable and accelerate high-fidelity, large-scale predictive simulation, preserving key capabilities.
We focus on two particularly valuable capabilities: efficient computation accelerator-based exascale platforms and exact differentiation for sensitivity analysis.
Existing approaches to computational thermochemistry provide, at most, only one such capability.
Herein, we expand on the utility of these capabilities and the challenges they present.

The evaluation of the thermochemistry often adds significantly to the cost of simulations of reacting flows.
This cost primarily follows three sources.
The first is an increase in the number of transported variables with respect to simulations of inert flow; the mixture composition can be described by up to hundreds of reacting scalars.
The second source is the evaluation of up to thousands of exponentials that appear in expressions representing reaction rates.
Last is a time-step restriction that increases the time of solution and stems from the imposed chemical time scale, over 1000~times smaller than that of the flow time scale~\cite{ref:MaasPope1992}.
These effects compound, and reacting flow simulations can be prohibitively expensive.
Compute accelerators (XPUs), including graphic processing units (GPUs) and accelerated processing units (APUs, or superchips), can mitigate this cost.

Our approach complements established thermochemistry software, such as dedicated libraries like  Cantera~\cite{ref:Cantera}, Chemkin~\cite{ref:Kee1996}, and TChem~\cite{ref:TChem}, which provide broad modeling capabilities but are limited to CPU hardware.
While serving the community well, these are not easily optimized for high-performance simulation, especially on accelerator hardware.
This is because the compiled libraries are standalone and live in a different executable than the flow solver.
Because of their breadth and structure, adapting such libraries to GPUs is an extensive task that risks conflicting with the goals and success of these tools.
Writing new thermochemistry software exclusively for large-scale flow simulations on accelerators risks sacrificing the modeling and symbolic capabilities that libraries provide. It also might face new roadblocks as compute hardware continues to evolve.
We develop and implement an alternative approach to thermochemistry computation based on symbolic abstractions.
From these, performant architecture-specific code is generated.

Our approach importantly enables computation of solution sensitivities to thermochemical variables and parameters for control~\cite{ref:Capecelatro2019} and uncertainty quantification~\cite{ref:Braman2015,ref:Zhao2019}.
One must carry corresponding derivatives to obtain these sensitivities.
The derivatives of the chemical source terms with respect to the composition vector, the chemical Jacobians, are the basis for various model reduction methods~\cite{ref:MaasPope1992,ref:LamGoussis1989,ref:ValoraniPaolucci2009} and state-of-the-art time integration~\cite{ref:Hindmarsh2005,ref:MacArtMueller2016,ref:Rao2025}, addressing numerical stiffness.

The implementation of analytical Jacobians challenges efficiency, and numerical approximations via finite differences require ad~hoc tuning of the perturbation size to meet accuracy requirements~\cite{ref:Niemeyer2017}.
Automatic differentiation (AD) is an attractive alternative because it provides the exact derivatives of a function without having to implement them manually in the source code~\cite{ref:Grienwank2003}.
In practice, AD represents functions as graphs and calculates corresponding derivatives via the chain rule as it traverses them. 
AD libraries and thermochemistry libraries typically belong to disparate compilation units, so the necessary graphs cannot be easily constructed.
The approach herein is AD-compatible via the same abstractions that enable accelerated, device-offloaded computation.

\begin{figure}[t]
  \begin{center}
    \includegraphics[scale=1,page=1]{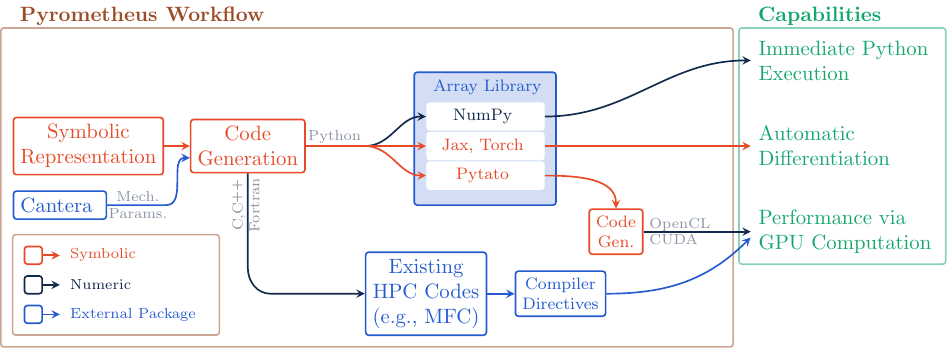}
    \caption{
        Pyrometheus generates code for combustion thermochemistry that can be transformed to meet the user's goals.
        Based on a symbolic representation of the formulation (\cref{sec:formulation}), it can produce Python, C, and Fortran code.
        The generated Python code can be executed with NumPy, automatically differentiated with compatible libraries like JAX, transformed to OpenCL or CUDA, or offloaded with directives like OpenACC or OpenMP for accelerator execution.
    }
    \label{fig:pyro_workflow}
  \end{center}
\end{figure}

\subsection{Principal Contribution}

This work's principal contribution is a hierarchy of abstractions that enable portable and performant computation for combustion thermochemistry.
It facilitates symbolic analysis and GPU execution for performance while minimizing barriers to entry or user intervention.
The implementation, called Pyrometheus, is open-source and permissively licensed.
The workflow and capabilities are illustrated in \cref{fig:pyro_workflow}.
The approach uses symbolic programming, presented in \cref{subsec:symbolic}, to abstract the mathematical operations of combustion thermochemistry away from the details of the specific simulation data on which they are performed.
The symbolic representation generates common workhorse languages for scientific computing, including Python, \Cpp{}, and Fortran via code generation mappings introduced in \cref{subsec:codegen}.
From the generated code, there are two paths to computation.
The first retains a Python control environment, from which the code can be immediately executed, automatically differentiated, or offloaded to accelerator hardware.
We achieve this environment via intermediate representations at progressively lower levels of abstraction (such as data flow graphs), which are obtained by staged addition of detail such as array semantics and loop patterns (\cref{subsec:ir_gpus}).
These transformations are demonstrated in \cref{sec:results}.
The second path to computation supports \Cpp{} or Fortran, which have their idiosyncratic control environments and offloading strategies.
This approach puts the computation of the thermochemistry at the same level of abstraction as the flow.
This path to computation is demonstrated in \cref{sec:chemfc}.

\subsection{Related Work \& Outline}

In some sense, Pyrometheus overcomes the challenges associated with analyzing and porting Cantera (and related libraries) to GPUs.
Cantera imposes strong assumptions on inputs to source-term computation: they are assumed to be pointwise quantities stored in memory as double-precision floats.
These assumptions further condition subsequent thermochemistry calculations (implemented as classes, deeply-nested loops, and conditionals), performance, and capabilities.
Principally, computational graphs cannot be obtained from such libraries due to the imposed data types.
These assumptions do not constrain Prometheus-generated code, so it provides exact derivatives and GPU execution via computational graph analysis, which are added capabilities over Cantera. Note that, as shown in \cref{fig:pyro_workflow}, Pyrometheus still relies on Cantera to obtain mechanism-specific data (such as Arrhenius parameters) at code generation. 

Pyrometheus is related to the pyJac library of \citet{ref:Niemeyer2017}, which generates CUDA code for GPU thermochemistry evaluation from Cantera. Particularly, pyJac implements analytical Jacobians for GPU evaluation, leveraging sparsity and memory access patterns~\cite{ref:Curtis2018}. Similarly, the Pele suite generates \Cpp{} code for quasi-steady state (QSS) chemical source terms and their Jacobians~\cite{ref:Hassanaly2024}, but rely on a Pele-intrinsic portability layer to offload the computation to GPUs~\cite{ref:Henry2022}. 
Pyrometheus has the added advantage that it is not tied to machine-specific languages (e.g., CUDA or \Cpp{}) and is performance-portable across vendors (e.g., AMD, NVIDIA, and Intel accelerators): the same Pyrometheus-generated code can be executed on a CPU, or re-expressed to run on accelerators, which are demonstrated in \cref{sec:results}. 
Pyrometheus is also advantageous in that neither the thermochemistry engine (Cantera) nor the user needs to provide explicit Jacobian-generation code due to the incorporation of automatic differentiation (AD).
This aspect is related to the Arrhenius.jl, the Julia code of \citet{ref:Ji2021} that uses AD to calibrate combustion mechanisms.
However, Pyrometheus and Arrhenius.jl serve different goals.
Pyrometheus is oriented towards large-scale simulations, supporting external target languages and control environments.

This work is also related to the computational design for chemical source terms of \citet{ref:Barwey2022}.
In their approach, the equations of combustion thermochemistry are cast in matrix form to mirror the evaluation of artificial neural networks (ANNs).
In this way, GPU-optimized ANN libraries can be used to compute the source terms.
However, their formulation is limited to computation on GPUs without derivatives, a limitation we overcome with Pyrometheus.

More broadly, our approach is related to the Pystella package~\cite{ref:Weiner2021}, a Python-based partial-differential equation (PDE) solver for astrophysics applications that achieves performance on GPUs through code generation and abstractions. Our approach also shares design principles with the work of Kulkarni~\cite{ref:Kulkarni2023}, where computational concerns are separated to achieve near-roofline GPU performance for finite-element methods.
However, Pyrometheus focuses on chemical source terms, whereas Pystella and Kulkarni target numerical discretization of differential operators. 

The paper is organized as follows.
The novel computational approach is introduced in \cref{sec:design}, and code correctness is confirmed in \cref{sec:verif}.  
Example implementations of thermochemistry in Python and quantitative results are presented in \cref{sec:results}.
Incorporation into a high-performant flow solver called MFC is demonstrated in \cref{sec:chemfc}.
\Cref{sec:impact} summarizes the principal features and discusses potential of extensions.


\section{Computational Approach}\label{sec:design}

\subsection{Goal \& Choice of Computational Paradigm}

Here, we present computational representations of combustion thermochemistry that meets the stated goals of \cref{sec:introduction}: to run at credibly high performance on both CPUs and GPUs and to be automatically differentiable. We first review the OpenCL--CUDA computational paradigm that bridges CPUs and GPUs, and provide a brief summary of automatic differentiation. This is followed by the development of our approach. 

To achieve portable performance, we invoke the OpenCL--CUDA model of computation, summarized next.
A computational platform is a hierarchy of resources: a host that connects to compute devices (either CPUs or GPUs), each composed of multiple compute units with processing elements.
The host controls program execution, and submits computation to the devices.
Languages like CUDA and OpenCL provide constructs to leverage the topology of the device and parallelize a computation across work groups that map onto the compute units. At a high level, the programmer interacts with data through arrays: sequences of objects arranged in memory. Our representations are designed to work within this computational paradigm with the added constrain that they must be compatible with AD libraries.  

AD libraries re-express a computation as a graph. As the graph is traversed, the chain rule is applied to yield the computation's derivatives with respect to its inputs. To obtain the graph from given code, AD libraries implement their own types and redefine a language's primitive operations. This approach is seemingly unrelated---or even at odds---with OpenCL--CUDA model of computation. However, we introduce computational representations that tie these goals through abstraction and subsequent progressive addition of detail. 

\subsection{A Symbolic Computational Representation of Combustion Thermochemistry}\label{subsec:symbolic}

The thermochemistry formulation that Pyrometheus implements is provided in full in \cref{sec:formulation}. Here, we focus on selected key expressions to guide development. In combustion simulations with detailed chemistry, one must evaluate forward reaction rates. For an (arbitrary) elementary binary reaction, the forward reaction rate is
\begin{equation}\label{eq:rxn_rate}
    R_{f} = k(T)C_{m}C_{n},
\end{equation}
where $T$ is the temperature, $C_{m}$ and $C_{n}$ are molar concentrations of the reactants, and
\begin{equation}\label{eq:arrhenius}
    k(T) = AT^{b}\exp\pp{ -\frac{\theta_{a}}{T} }
\end{equation}
is the rate coefficient, with parameters $\{ A,\, b,\, \theta_{a} \}$.
Regardless of the application, \cref{eq:rxn_rate} must be repeatedly evaluated: across reactions $j = 1,\, \dots,\, M$, but also spatial points (or notional particles), time steps, and possibly parametric realizations for uncertainty quantification.
However, the computation is local (without spatial gradients), and there are no loop-carried dependencies, so it is suitable for an OpenCL and CUDA style of parallelism.    

We seek a computational representation of \cref{eq:rxn_rate} and \cref{eq:arrhenius} that incorporates the low-level constructs that enable parallelization across multiple logical units on CPUs and GPUs.
One could write OpenCL or CUDA code, but this would preclude AD and could also conflict with the parallelization strategies used by existing flow solvers.
However, these challenges are computational, independent of \cref{eq:rxn_rate} and \cref{eq:arrhenius} (and the broader formulation of \Cref{sec:formulation}).
We exploit this property by elevating the computation to a symbolic level that more closely matches the formulation.
At this symbolic level, the representation expresses only mathematical operations, and distinctions between specific cases (e.g., GPU execution versus AD-enabled time marching) vanish. 

At the highest level of abstraction, which is closest to the user, in the Pyrometheus workflow (\cref{fig:pyro_workflow}), the computational representation is symbolic: not yet with numerical values of the thermochemistry computation.
For this reason, we implement algebraic routines in Python due of its flexible data model.
It provides a straightforward means of emulating numeric types by overriding arithmetic methods for user-defined objects. 

\begin{figure}[ht]
    \begin{center}
        \includegraphics[scale=1,page=1]{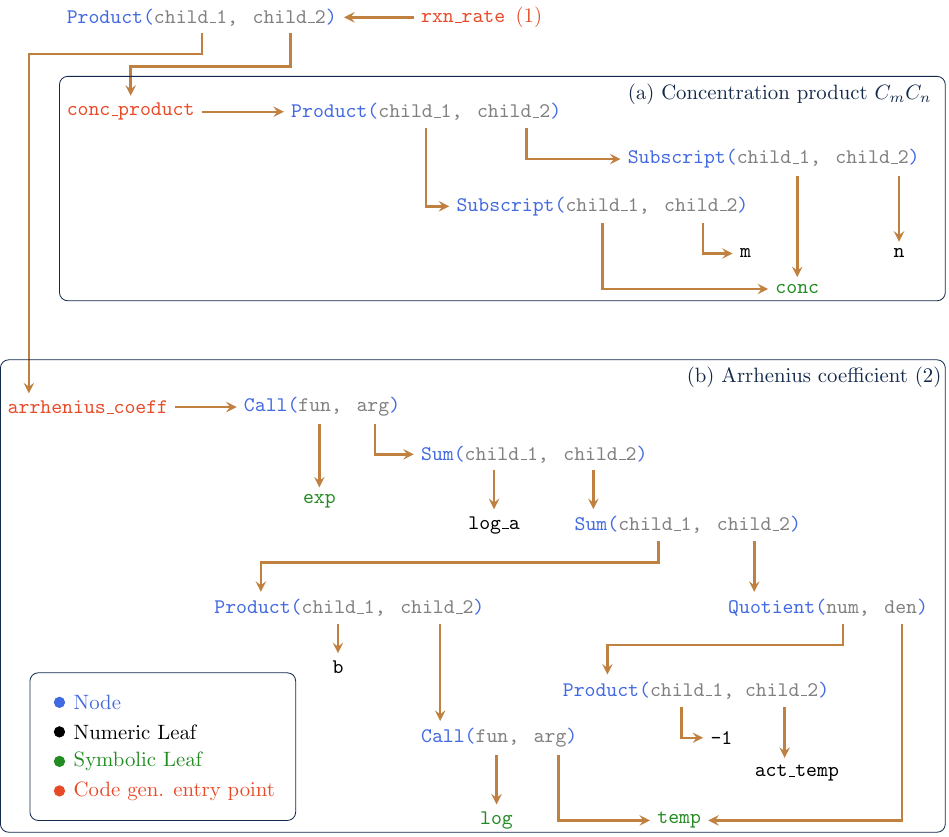}
        \caption{
            Expression trees for the reaction rate \cref{eq:rxn_rate}: (a) the $C_m C_n$ term in \cref{eq:rxn_rate}; and (b) Arrhenius coefficient \cref{eq:arrhenius}.
            Nodes (blue) are operations defined in \cref{l:symbolic_expr}.
            Leaves (green) are variables defined in \cref{l:symbolic_var}; nonlinear functions are treated as leaves, mapped to specific implementations at code generation.
            Code generation entry points (in orange) are nodes from which expressions in target languages are formed.
        }
        \label{fig:pyro_graph}
    \end{center}
\end{figure}

\Cref{fig:pyro_graph} shows the necessary algebra to arrange expressions into trees.
Trees are useful data structures here because they can be efficiently manipulated via traversal, enabling multi-language code generation.
\Cref{subsec:codegen} describes this more fully.
More complex expressions can also be created by adding nodes to the tree.
For example, the tree in \cref{fig:pyro_graph}~(b) can be extended from $k$ to $R_f$ \cref{eq:rxn_rate} by adding the tree for the concentration product in \cref{fig:pyro_graph}~(a).

We formalize expression trees through an algebraic system.
The following development is pragmatic, combining precise definitions and examples; this is sufficient to achieve our goals.
Because we seek to build computational objects that represent thermochemistry, we rely on code listings rather than equations.
This approach is as rigorous as an explicit mathematical treatment yet more efficient.
Furthermore, this choice dispels ambiguity that can arise between computational objects and the mathematical terms of the formulation they are meant to represent.
We present just the key portions of the code; the full development can be found in \url{https://github.com/pyrometheus/mini-pyrometheus}.

Expression trees, such as in \cref{fig:pyro_graph}, are defined via their nodes, each of which is itself an expression.
This definition is shown in \cref{l:symbolic_expr}, and, while still (somewhat) abstract, it is the base for more concrete expressions such as sums and products.
The \lstinline|children| (line 4) attribute represents the nodes and solely defines an expression as a computational object.
For example, the \lstinline|rxn_rate| expression of \cref{fig:pyro_graph} has two children: one that represents the concentration product $C_m C_n$ and one that represents the Arrhenius coefficient $k(T)$ \cref{eq:arrhenius}.
In lines~9--13, the \lstinline|Expression| class overloads (redefines) the built-in algebraic Python methods.
Each overloaded method returns a concrete expression type; for example, multiplying two objects of type \lstinline|Expression| results in a \lstinline|Product| object.
The overloaded methods, along with specific node (expression) types, are key to forming the computational expressions that represent thermochemistry.

\begin{lstlisting}[
float,
floatplacement=ht,
language=Python,
basicstyle={\footnotesize\ttfamily\color{myBlue}},
linerange={1-13},
label={l:symbolic_expr}, 
caption={
    Symbolic computational expressions underpin the symbolic representation.
    Concrete expressions such as \lstinline|Sum| and \lstinline|Product| have a \lstinline|mapper_method| tag to guide the code generation of \cref{l:codegen_python}.
    The release makes use of an extended version of this, based on Pymbolic~\cite{ref:pymbolic}.
}
]
class Expression:
    def __init__(self, children):
        self.children = children
    def __add__(self, other):
        return Sum((self, other))
    def __mul__(self, other):
        return Product((self, other))

class Sum(Expression):
    mapper_method = 'map_sum'

class Product(Expression):
    mapper_method = 'map_product'
\end{lstlisting}

The simplest type of node is a leaf: it does not have children.
In the implementation of \cref{l:symbolic_var}, a leaf object is called a \lstinline|Variable|, defined solely by its name (a string).
As shown in \cref{fig:pyro_graph}, state variables such as temperature and concentration are leaves.
Correspondingly, we create two variables in \cref{l:symbolic_var} lines~8 and~9.
Subsequent expression trees in the development are built on these two objects for $T$ and $\vec{C}$.
The release makes use of this same step to initialize the formation of thermochemistry expressions.
We note that, at this abstraction level, function names are also of type \lstinline|Variable|.
Distinguishing between function names and variables is conducted at the code generation stage of \cref{subsec:codegen}.

\begin{lstlisting}[
float,
floatplacement=!ht,
language=Python,
basicstyle={\footnotesize\ttfamily\color{myBlue}},
linerange={15-25},
label={l:symbolic_var}, 
caption={
    A variable is a leaf in an expression tree, defined solely by its name (a string).
    For \cref{eq:rxn_rate} and \cref{eq:arrhenius}, temperature $T$ and species molar concentrations $\vec{C} = \{ C_{m},\,C_{n}\}$ are leaves, so defined in lines~6 and 7.}
]
class Variable(Expression):
    mapper_method = 'map_variable'
    def __init__(self, name):
        self.name = name

temp = Variable('temperature')
conc = Variable('concentration')
print(temp.name)
# >>> temperature
print(conc.name)
# >>> concentration
\end{lstlisting}

From variables, we build concrete expressions such as the sum and product objects of lines 9--13 in \cref{l:symbolic_expr}.
Because variables are also expressions, they inherit the overloaded algebraic methods of \cref{l:symbolic_expr}.
Two variables can be multiplied to form a \lstinline|Product| expression, so the $C_{m}C_{n}$ term in \cref{eq:rxn_rate} can be represented via the syntax of \cref{l:symbolic_math_ex}.
Once built, the corresponding expression tree of \cref{fig:pyro_graph}~(a) can be traversed by recursively accessing the children in \lstinline|conc_product|.
With this approach, complex expression trees (\cref{fig:pyro_graph}) are created with the same standard Python syntax (\cref{l:symbolic_math_ex}) that is used to manipulate numbers. 

\begin{lstlisting}[
float,
floatplacement=!ht,
language=Python,
basicstyle={\footnotesize\ttfamily\color{myBlue}},
linerange={27-48},
label={l:symbolic_math_ex}, 
caption={
    The expression tree for the concentration product $C_m C_n$ of \cref{eq:rxn_rate} and \cref{fig:pyro_graph}~(a), built from the symbolic variables in \cref{l:symbolic_var}.
    Lines~3--13 implement recursive traversal of the \lstinline|conc_product| tree, with the output shown in lines~16--22.
}
]
conc_product = conc[0] * conc[1]

def rec_example(expr):
    if isinstance(expr, Variable):
        print(f'\t Reached a leaf with name {expr.name}')
        return
    elif isinstance(expr, int):
        print(f'\t Reached a numeric leaf with value {expr}')
        return
    else:
        print(f'Visited node of type {type(expr)}')

    for c in expr.children: rec_example(c)

rec_example(conc_product)
# >>> Visited node of type <class 'symbolic.Product'>
# >>> Visited node of type <class 'symbolic.Subscript'>
# >>>      Reached a symbolic leaf with name concentration
# >>>      Reached a numeric leaf with value 0
# >>> Visited node of type <class 'symbolic.Subscript'>
# >>>      Reached a symbolic leaf with name concentration
# >>>      Reached a numeric leaf with value 1
\end{lstlisting}

To complete the symbolic representation of \cref{eq:rxn_rate}, we need to form the Arrhenius coefficient \cref{eq:arrhenius}.
For this, the parameters $\{A,\, b,\, \theta\}$ must be represented somehow.
We assume these parameters are numeric and encapsulated in a structure with a properly defined Python interface that retrieves them.
Thermochemistry libraries provide such capabilities; in particular, we use the Cantera library, which is open source and has an extensive Python interface. 

Expression tree assembly for $k$ \cref{eq:arrhenius}, using Cantera, is shown in \cref{l:symbolic_arrhenius}.
The full release implements the full formulation of \cref{sec:formulation} in this way, by combining symbolic \lstinline|Variables| and Cantera objects.
In this sense, our implementation is tied to Cantera or at least a library with a similar Python interface.
However, Cantera is comprehensive and provides additional capabilities, such as handling plasmas.
Extensions are discussed in \cref{sec:impact}.

\begin{lstlisting}[
float,
floatplacement=!ht,
language=Python,
basicstyle={\footnotesize\ttfamily\color{myBlue}},
linerange={10-32},
label={l:symbolic_arrhenius}, 
caption={
    The expression trees for the Arrhenius coefficient \cref{eq:arrhenius} and forward reaction rate \cref{eq:rxn_rate}.
    The trees combine symbolic variables (\cref{l:symbolic_var}) and numeric data for Arrhenius parameters, encapsulated in a Cantera \lstinline|Reaction| object.}
]
def arrhenius_expr(rxn: ct.Reaction, temp: Variable):
    # Nonlinear functions
    exp = Variable('exp')
    log = Variable('log')
    # Arrhenius parameters
    log_a = np.log(rxn.rate.pre_exponential_factor)
    b = rxn.rate.temperature_exponent
    act_temp = -1 * rxn.rate.activation_energy / ct.gas_constant
    # Construct the Arrhenius expression
    return exp(log_a + b * log(temp) + act_temp / temp)

rxn = ct.Reaction(
    # Reactants & Products for M + N -> P + Q
    {'m': 1, 'n': 1}, {'p': 1, 'q': 1},
    # Arrhenius coefficients, taken from Reaction 1, San Diego mech
    {'A': 35127309770106.477, 'b': -0.7, 'Ea': 8590 * ct.gas_constant})
k = arrhenius_expr(rxn, temp)
rxn_rate = k * conc_product

print(f'{isinstance(k, Expression)}')
# >>> True
print(f'{isinstance(rxn_rate, Expression)}')
# >>> True
\end{lstlisting}

\subsection{Code Generation via Translation of the Symbolic Representation}\label{subsec:codegen}

The next step towards computation is generating code from the symbolic representation.
Our approach, presented in what follows, is independent of the language of the generated code.
We illustrate the approach for Python; the generated code is designed to facilitate transformations for GPU execution \textit{and} be compatible AD tools.
The release provides Fortran and \Cpp{} support based on the same approach.

We follow a standard approach to code generation that takes expression trees as inputs and returns code in the target language as a string~\cite{ref:Dragon}.
The approach is embodied in \cref{l:codegen_python}.
As shown in \cref{fig:pyro_graph}, the expression trees \lstinline|conc_product| and \lstinline|k| are the entry point to code generation.
Broadly, these trees are selected as entry points because they are mechanism-specific.
The trees are traversed, and translation rules (mapper methods in \cref{l:codegen_python}) are recursively applied to each of their children (\cref{l:codegen_python} lines~20 and 25).
For each node, the mapper method creates code that is concatenated from smaller fragments that preserve the node's meaning and are grammatically correct in the target language. 

\Cref{l:codegen_fragments} shows the code for \lstinline|conc_product| and \lstinline|k|.
For the Arrhenius expression in Python, we assume there is a NumPy-like array library \lstinline|pyro_np| that computes the exponential element-wise on arrays.
In this sense, the array-based operations in the generated code are still abstract by design, and the generated code remains symbolic.
We discuss concrete choices for the array library and their implications in \cref{subsec:ir_gpus}. 

\begin{lstlisting}[
float,
floatplacement=!ht,
language=Python,
basicstyle={\footnotesize\ttfamily\color{myBlue}},
linerange={1-30},
label={l:codegen_python}, 
caption={
    A code generation mapper represents the rules used to generate code from the symbolic representation.
    Each mapper method (\lstinline|map_*|) corresponds to a rule that preserves the meaning of the expression being translated.
    The methods are recursively applied via \lstinline|rec|, concatenating code fragments into a corresponding target-language expression.
}
]
class CodeGenerationMapper:
    prec = {'var': 0, 'call': 1, 'sum': 2, 'mul': 3, 'div': 4, 'sub': 5}

    def parenthesize(self, expr_str, prec_expr, prec):
        if prec and prec > prec_expr:
            return f'({expr_str})'  # parenthesize result
        else:
            return expr_str

    def rec(self, expr, *args):
        if args:
            return getattr(self, expr.mapper_method)(expr, *args)
        else:
            return getattr(self, expr.mapper_method)(expr, None)

    def map_variable(self, expr, precedent): return expr.name

    def map_sum(self, expr, prec):
        return self.parenthesize(
            ' + '.join([self.rec(c, self.prec['sum']) for c in expr.children]),
            self.prec['sum'], prec)

    def map_product(self, expr, prec):
        return self.parenthesize(
            ' * '.join([self.rec(c, self.prec['mul']) for c in expr.children]),
            self.prec['mul'], prec)
\end{lstlisting}

\begin{lstlisting}[
float,
floatplacement=!ht,
language=Python,
basicstyle={\footnotesize\ttfamily\color{myBlue}},
linerange={7-32},
label={l:codegen_fragments}, 
caption={
    Code generation as applied to the symbolic concentration product and Arrhenius coefficient.
    When applied recursively to a symbolic \lstinline|Expression|, the code generation mapper \lstinline|cgm| emits a code fragment as a string.
}
]
import cantera as ct
import numpy as np

def arrhenius_expr(rxn: ct.Reaction, temp: Variable):
    # Nonlinear functions
    exp = Variable('exp')
    log = Variable('log')
    # Arrhenius parameters
    log_a = np.log(rxn.rate.pre_exponential_factor)
    b = rxn.rate.temperature_exponent
    act_temp = -1 * rxn.rate.activation_energy / ct.gas_constant
    # Construct the Arrhenius expression
    return exp(log_a + b * log(temp) + act_temp / temp)

rxn = ct.Reaction(
    # Reactants & Products for M + N -> P + Q
    {'m': 1, 'n': 1}, {'p': 1, 'q': 1},
    # Arrhenius coefficients, taken from Reaction 1, San Diego mech
    {'A': 35127309770106.477, 'b': -0.7, 'Ea': 8590 * ct.gas_constant})
k = arrhenius_expr(rxn, temp)
rxn_rate = k * conc_product

print(f'{isinstance(k, Expression)}')
# >>> True
print(f'{isinstance(rxn_rate, Expression)}')
# >>> True
\end{lstlisting}

The code generation procedure is embedded in a code template, which is static text in the target language with functionality independent of the formulation's details.
This is the same technology used to generate web pages.
The Python template is shown in \cref{l:python_template}.
The key design choice is that the generated code depends on the unspecified \lstinline|pyro_np| for processing its array-based expressions.
This separates computational concerns: the generated code represents the formulation, and the burden of array-based computation is placed on an external tool.
The generated code can already be used to compute thermochemistry via NumPy.
Furthermore, because NumPy processes arrays of arbitrary dimensions, the same generated code can simulate auto-ignition in homogeneous reactors and flames without looping over a spatial domain.
This property is showcased in \cref{sec:results} using NumPy as the array library. 

\begin{lstlisting}[
float,
floatplacement=!ht,
%language=Python,
basicstyle={\footnotesize\ttfamily\color{myBlue}},
linerange={1-20},
label={l:python_template}, 
caption={
    Python template for thermochemistry code to compute \cref{eq:rxn_rate}.
    The syntax \lstinline|\{ \}|, where code generation is embedded, stands for substitution.
    The \lstinline|pyro_np| input to class construction represents a yet-unspecified library for array computation.
    The release implements a template for the full formulation of \cref{sec:formulation}.
}
]
from mako.template import Template

code_tpl = Template("""
class Thermochemistry:
    def __init__(self, pyro_np):
        self.pyro_np = pyro_np

    def get_rxn_rate(self, temperature, concentration):
        k = ${cgm.rec(arrhenius_expr(rxn, Variable("temperature")))}
        conc_product = ${cgm.rec(conc_product)}
        return k * conc_product
""", strict_undefined=True)

code_str = code_tpl.render(
    Variable=Variable,
    arrhenius_expr=arrhenius_expr,
    rxn=rxn,
    cgm=python_cgm,
    conc_product=conc_product
)
\end{lstlisting}

Upon rendering, shown in lines 14--22 of \cref{l:python_template}, code fragments are generated, concatenated, and inserted into the template.
The generated code \lstinline|code_str| in \cref{l:generated_python} can run on CPUs.
Code correctness is established at this stage; the procedure is discussed in detail in \cref{sec:verif}.
As stated, the generated code can already compute the thermochemistry with NumPy as its array processing library.
The necessary extensions for GPU execution and automatic differentiation are discussed next. 

\begin{lstlisting}[
float,
floatplacement=!ht,
language=Python,
basicstyle={\footnotesize\ttfamily\color{myBlue}},
label={l:generated_python}, 
caption={Generated Python code for \cref{eq:rxn_rate}. The array library \lstinline|pyro_np|, yet unspecified, assigns meaning to the array-valued expressions that compose the code.}
]
class Thermochemistry:
    def __init__(self, pyro_np):
        self.pyro_np = pyro_np
    def get_rxn_rate(self, temperature, concentration):
        k = self.pyro_np.exp(
            31.19 - 0.7 * self.pyro_np.log(temperature) - 8590.0 / temperature)
        conc_product = concentration[0] * concentration[1]
        return k * conc_product
\end{lstlisting}

\subsection{Intermediate Representations and the Path to AD \& GPU Computation in Python}\label{subsec:ir_gpus}

Python is a scripting language with relatively low proficiency barriers with respect to, say, \Cpp{}.
Yet, it provides stable and comprehensive access to HPC tools for communication (mpi4py), parallel IO (h5py), and device access (PyOpenCL and PyCUDA)~\cite{ref:Barba2021}.
It is advantageous to retain control of computation in such environments~\cite{ref:Slotnick2014,ref:Betcke2021}.
Next, we develop a framework to achieve AD and HPC of the thermochemistry in Python. 

The proposed framework is based on data flow graphs (DFG) and the transformations that give rise to the capabilities of \cref{fig:pyro_workflow}. The edges and nodes of our DFG represent arrays and the operations of the thermochemistry.
The DFG extends the expression trees of \cref{fig:pyro_graph} by accounting for numerical data, such as actual values, array shapes, and data types.
We will show how to extract DFG from our generated code.
However, users will still interact with the generated Python code, so the DFG and subsequent transformations are intermediate representations (IR) of the thermochemistry. 

The generated code of \cref{l:generated_python} assumes only that array operations are conducted by an array processing library that resembles NumPy in its syntax, e.g., \lstinline|pyro_np.exp| in line~5 of \cref{l:generated_python}.
No further assumptions are made about the numerical data it ultimately processes.
This design choice reflects the separation of computational concerns: the generated code provides the expressions of the thermochemistry, and the array library assigns them the necessary meaning to achieve a specific capability.
The meaning of an array expression, such as 
\begin{equation}
    \texttt{concentraction[0] * concentration[1]}\notag
\end{equation}
in line~7 of \cref{l:generated_python}, can be extended from simply the operation it expresses to include information about, e.g., derivatives, memory concerns, and devices. Thus, as shown \cref{fig:pyro_workflow}, the choice of array library determines the specific capability of the generated code.

Our full release makes use of comprehensive array libraries such as JAX for AD and Pytato for GPU offloading.
Once the thermochemistry code is generated, it can be paired with any of these libraries; concrete examples are presented in \cref{sec:results}.
However, it is useful to understand the underlying constructs and transformations that these libraries provide, as it can lead to extended capabilities of the generated code.
For this, we develop two custom array libraries next to achieve AD and GPU offloading with the generated code of \cref{l:generated_python}.
We first introduce the methods to extract the DFG and apply it to AD.
Then, to offload the computation to GPUs, we present an array library underpinned by a sequence of code transformations.


\subsubsection{Custom Arrays to Showcase Automatic Differentiation}\label{subsub:ad_arrays}

Automatic differentiation allows us to obtain
\begin{equation}\label{eq:rate_gradient}
    \mathcal{G} \equiv \Bigg\{ \frac{\partial R_{f}}{\partial T},\, \frac{\partial R_{f}}{\partial C_{m}},\, \frac{\partial R_{f}}{\partial C_{n}} \Bigg\}
\end{equation}
without approximation nor code generation for approximating derivatives.
Obtaining analytical expressions for the derivatives in \cref{eq:rate_gradient} is trivial, but it is more challenging for the formulation of \cref{sec:formulation}, so AD is an attractive alternative.
In AD, the computation is expressed as a DFG, and derivatives are obtained by applying the chain rule along the path from output $R_{f}$ to each input $\{ T,\, C_{m},\, C_{n} \}$.
The gradient of each node along the path is computed from stored values and propagated to its children.  

The DFG can be constructed when we evaluate the function \lstinline|get_rxn_rate| of \cref{l:generated_python} line~4.
For this, we create a custom array type, shown in \cref{l:ad_array}, containing numerical data as well as an attribute to track computational dependencies.
As in \cref{subsec:symbolic}, this custom array has redefined algebraic methods.
These methods process numerical arrays and provide the functionality to compute the gradient of the corresponding operation (as \lstinline|grad_fn|).
The custom array types and their methods compose the custom \lstinline|adiff_np| array library that illustrates the compatibility of our approach with JAX and PyTorch.

\begin{lstlisting}[
float,
floatplacement=!ht,
language=Python,
basicstyle={\footnotesize\ttfamily\color{myBlue}},
linerange={3-24},
label={l:ad_array}, 
caption={
    Arrays for automatic differentiation, illustrative of AD arrays in JAX and tensors in PyTorch.
    The arrays combine numeric values with symbolic capabilities (via the \lstinline|children| attribute).
    Chain rule expressions are implemented as \lstinline|grad_fn| for concrete DFG node types such as \lstinline|AutodiffVariable| and \lstinline|AutodiffProduct|.
}
]
class AutodiffArray:
    def __init__(self, values: list, children, name=None):
        self.values = np.array(values, dtype=np.float64)
        self.children = children
        self.name = name
    def __mul__(self, other):
        return AutodiffProduct(self.values * other.values, children=[self, other])
    def gradient(self,):
        ad_walker = AutodiffWalker()
        return ad_walker.compute_gradient(self)

class AutodiffVariable(AutodiffArray):
    def __init__(self, values, name):
        self.values = values
        self.name = name
        self.children = []
    def grad_fn(self, grad):
        return np.zeros_like(grad)

class AutodiffProduct(AutodiffArray):
    def grad_fn(self, grad):
        return (grad * self.children[1].values, grad * self.children[0].values)
\end{lstlisting}

The generated code is paired with the custom array library in \cref{l:ad_ex}, which further shows computation on the numerical $\{T,\, C_{m},\, C_{n} \}$.
As \lstinline|get_rxn_rate| processes the numerical values of $R_{f}$ for given $\{ T,\, C_{m},\, C_{n} \}$, it also assembles the DFG.
Importantly, each node in the DFG has numerical values associated with its children, and these will be exploited to compute \cref{eq:rate_gradient}.
The AD process is implemented in \cref{l:ad_mapper} as a recursion that propagates gradients along each path from root (output) to leaves (inputs).
The entry point to this recursion is the \lstinline|compute_gradient| method of line~3.
As shown in line~8 of \cref{l:ad_ex}, the DFG provides an interface to the AD entry point, substantially simplifying the control sequence for users.
This approach resembles the interfaces of AD libraries such as JAX and PyTorch.

\begin{lstlisting}[
float,
floatplacement=!ht,
language=Python,
basicstyle={\footnotesize\ttfamily\color{myBlue}},
linerange={26-45},
label={l:ad_ex}, 
caption={
    Automatic differentiation of the generated Python code in \cref{l:generated_python}.
    Temperature and concentrations are represented with the array type of \cref{l:ad_array}.
    The \lstinline|gradient| method in line~8 calls the recursive algorithm to compute derivatives exactly.
}
]
class AutodiffWalker:
    def compute_gradient(self, ary):
        prec_grad = np.ones_like(ary.values)
        self.gradients = {}
        self.rec(ary, prec_grad)
        return self.gradients

    def rec(self, ary, prec_grad):
        if isinstance(ary, AutodiffVariable):
            if ary.name in self.gradients:
                self.gradients[ary.name] += prec_grad
            else:
                self.gradients[ary.name] = prec_grad
        else:
            self.propagate_gradients(ary, prec_grad)

    def propagate_gradients(self, ary, prec_grad):
        for c, g in zip(ary.children, ary.grad_fn(prec_grad)):
            if isinstance(c, AutodiffArray):
                self.rec(c, g)
\end{lstlisting}

\begin{lstlisting}[
float,
floatplacement=!ht,
language=Python,
basicstyle={\footnotesize\ttfamily\color{myBlue}},
linerange={47-54},
label={l:ad_mapper}, 
caption={Automatic differentiation: recursive propagation of derivatives along the DFG. The algorithm is called from the DFG via the \lstinline|gradient| method.}
]
temp_np = np.linspace(1000, 2100, 10, endpoint=False)
conc_np = np.stack((0.2 * np.ones(10), 0.8 * np.ones(10)))

pyro_gas = Thermochemistry(pyro_np=adiff_np)
temp_ad = autodiff.AutodiffVariable(temp_np, name='temperature')
conc_ad = autodiff.AutodiffVariable(conc_np, name='concentration')
rxn_rate = pyro_gas.get_rxn_rate(temp_ad, conc_ad)
g = rxn_rate.gradient()
\end{lstlisting}

Because the AD process is recursive, it is important to maintain data-independent control flow in the computation.
This condition is violated when the temperature is inverted from, for example, the internal energy, as is common in compressible flow solvers.
As shown in \cref{subsec:pyro_ad}, comprehensive libraries like JAX provide the necessary tools to handle data-dependent control flow.

\subsubsection{Custom Arrays with Underlying Transformations to Enable GPU Computing in Python}\label{subsub:lazy_arrays}

We now focus on offloading the computation of thermochemistry to GPUs, which is demonstrated via the CUDA and OpenCL computational models.
Here, we present the transformations for obtaining the necessary CUDA code from the generated Python code.
These transformations are realized as an array library, so control of the computation remains in Python.
The transformations are purposefully open and easily reached, so they can be continuously improved if needed.

The array types that compose the library for GPU offloading are shown in \cref{l:lazy_arrays}.
These arrays illustrate the comprehensive Pytato library, which is used in the demonstrations of \cref{sec:results} for the full thermochemistry formulation.
Note that, in contrast with \cref{subsub:ad_arrays}, we forgo the immediate specification of numerical values for the arrays (but retain the shape attribute).
This is to avoid the costly computation of the thermochemistry across large datasets on a host CPU as the DFG is constructed.
The deferred binding of the arrays to their data is known as lazy execution, so the custom array library is called \lstinline|lazy_np|.
Lazy evaluation is advantageous in heterogeneous computing as the data transfer from the host (CPU) to the compute device (GPU) can be the most expensive part of a workflow.

We identify two types of symbolic arrays: input placeholders (\cref{l:lazy_arrays}, line~8) and output array expressions (\cref{l:lazy_arrays}, line~13).
Placeholders contain only shape information, without numerical values of the thermochemistry.
The output array expression is the DFG and the shape of the resulting array.
In this demonstration, we determine output shape through NumPy broadcasting rules (\cref{l:lazy_arrays}, line~3), described in detail elsewhere~\cite{ref:numpy2011}. 

\begin{lstlisting}[
float,
floatplacement=!ht,
language=Python,
basicstyle={\footnotesize\ttfamily\color{myBlue}},
linerange={1-28},
label={l:lazy_arrays}, 
caption={
    Deferred-evaluation (lazy) arrays to offload the computation to GPUs.
    The \lstinline|compile| method executes a sequence of Loopy transformations to derive CUDA code from the DFG.
    The \lstinline|assemble_cuda| method (line~20) is shown in \cref{l:assemble_cuda}.}
]
class LazyArray:
    def __mul__(self, other):
        output_shape = np.broadcast_shapes(self.shape, other.shape)
        return ArrayExpression(expr=Product((self, other)),
                               shape=output_shape)
    # def __add__(self, other): ...

class Placeholder(LazyArray):
    def __init__(self, name, shape):
        self.expr = Variable(name)
        self.shape = shape

class ArrayExpression(LazyArray):
    def __init__(self, expr, shape):
        self.expr = expr
        self.shape = shape
        self.cuda_prg = None
    def compile(self, knl_name, wg_size):
        self.wg_size = wg_size
        from minipyro.pyro_np.loopy import assemble_cuda
        self.cuda_prg, self.cuda_code = assemble_cuda(self, knl_name)
    def evaluate(self, *np_data):
        assert self.cuda_prg is not None
        # grid = ...
        # block = ...
        import pycuda.gpuarray as gpuarray
        dev_data = [gpuarray.to_gpu(a) for a in np_data]
        self.cuda_prg(*dev_data, grid=grid, block=block)
\end{lstlisting}

The DFG expresses algebraic relations between arrays, not the loops that execute these, data layout, and potential parallelism.
For this, we make use of the Loopy paradigm~\cite{ref:Kloeckner2014}, which expresses a computation in terms of a Loopy kernel: loop bounds and a list of instructions.
The instructions express manipulation of array data, with memory access parameterized by loop variables that vary across corresponding bounds.
From the Loopy representation, we can generate OpenCL or CUDA code.
Complementary tools like PyOpenCL and PyCUDA embed execution in a Python control environment.

Loopy kernels can be generated from the DFG in the same way the Python code is generated from expression trees.
This code generation step is shown in \cref{l:assemble_cuda}.
We embed the generation of Loopy kernels as an intermediate step in the numerical evaluation of array-based expressions.
Subsequent code generation maps the Loopy kernel to CUDA code, which can be compiled and invoked through PyCUDA, retaining control of the computation in Python.

\begin{lstlisting}[
float,
floatplacement=!ht,
language=Python,
basicstyle={\footnotesize\ttfamily\color{myBlue}},
linerange={33-50},
label={l:loopy_knl}, 
caption={Loopy kernel generated from the DFG. This original kernel does not implement parallelization of the computation across work groups, which requires transformation \cref{eq:lp_split} of its indices.}
]
KERNEL: get_rxn_rate
---------------------------------------------------------------------------
ARGUMENTS:
concentration: type: np:dtype('float64'), shape: (2, 1024, 1024), ...
rxn_rate: type: np:dtype('float64'), shape: (1024, 1024), ...
temperature: type: np:dtype('float64'), shape: (1024, 1024), ...
---------------------------------------------------------------------------
DOMAINS:
{ [i0, i1] : 0 <= i0 <= 1023 and 0 <= i1 <= 1023 }
---------------------------------------------------------------------------
INAME TAGS:
i0: None
i1: None
---------------------------------------------------------------------------
INSTRUCTIONS:
for i1, i0
    rxn_rate[i0, i1] = ...  {id=insn}
end i1, i0
\end{lstlisting}

The original Loopy kernel obtained from the DFG, shown in \cref{l:loopy_knl}, does not implement parallelization.
In the original kernel, loop variables span their entire domain, so the corresponding CUDA code will sequentially execute the instructions for each element in the array, which is inefficient.
Efficient execution on GPUs is achieved by splitting loop-variable bounds into work groups that map onto individual computational units within the domain.
This transformation is~\cite{ref:Kloeckner2014}
\begin{equation}\label{eq:lp_split}
    k = i + g\,j,
\end{equation}
where $k$ is the original index, $i$ the inner (work-group local) index, $g$ the work-group size, and $j$ the outer (work-group global) index.
So, each work group executes part of the computation, with all groups executing simultaneously.

\begin{lstlisting}[
float,
floatplacement=!ht,
language=Python,
basicstyle={\footnotesize\ttfamily\color{myBlue}},
linerange={1-30},
label={l:assemble_cuda}, 
caption={Transformation of the DFG for \cref{eq:rxn_rate} into CUDA code, via Loopy. The intermediate Loopy kernel is transformed to account for parallelization via \cref{eq:lp_split}.}
]
def assemble_cuda(ary, knl_name):
    dim = len(ary.shape)
    idx_list = [f'i{i}' for i in range(dim)]

    lp_domains = ... # Create domains for loop indices

    lp_mapper = LoopyMapper()  # A codegen mapper for DFG -> Loopy instructions
    idx_tuple = ', '.join(idx_list)

    # Render a Loopy template
    lp_tpl = Template(
    """
    rxn_rate[${idx_tuple}] = ${expr}
    """, strict_undefined=True)
    lp_instructions = lp_tpl.render(idx_tuple=idx_tuple, expr=lp_mapper.rec(ary))

    import loopy as lp
    lp_knl = lp.make_kernel(lp_domains, lp_instructions, name=knl_name)
    ...
    # Parallelize by splitting loop domains
    for i in range(dim):
        lp_knl = lp.split_iname(lp_knl, f'i{i}', ary.wg_size,
                                outer_tag=f'g.{i}', inner_tag=f'l.{i}')
    # To CUDA
    lp_knl = lp_knl.copy(target=lp.CudaTarget())
    code_str = lp.generate_code_v2(lp_knl).device_code()

    from pycuda.compiler import SourceModule
    prg = SourceModule(code_str).get_function(knl_name)
    return prg, code_str
\end{lstlisting}

\begin{lstlisting}[
float,
floatplacement=!ht,
language=Python,
basicstyle={\footnotesize\ttfamily\color{myBlue}},
linerange={1-15},
label={l:gpu_exec}, 
caption={
    Python interface to the evaluation of \cref{eq:rxn_rate} on a GPU.
    A sequence of intermediate Loopy representations underlie the \lstinline|compile| method to generate CUDA code.
    This representation enables parallel computation in work groups of size $32$.
}
]
pyro_class = get_thermochem_class()
pyro_gas = pyro_class(lazy_np)

wg_size = 32
num_x = 32 * wg_size
temp = lazy_np.Placeholder(name='temperature', shape=(num_x, num_x))
conc = lazy_np.Placeholder(name='concentration', shape=(2, num_x, num_x))

rxn_rate = pyro_gas.get_rxn_rate(temp, conc)
rxn_rate.compile('get_rxn_rate', wg_size)

temp = 300 * np.ones((num_x, num_x))
conc = 0.5 * np.ones((2, num_x, num_x))
rate = np.zeros((num_x, num_x))
rxn_rate.evaluate(conc, rate, temp)
\end{lstlisting}

\section{Code Correctness}\label{sec:verif}

The generated code is then compared to those of Cantera, a well-tested mainstay library, to ensure its correctness.
Our tests cover the equation of state, species thermodynamic properties, and kinetic rates and their coefficients.
The equation of state tests ensures that pressure and temperature are correctly computed from known states.
Species thermodynamic properties are tested over the temperature range of validity for NASA polynomials.
Kinetic properties that depend on mixture composition are tested for various states along trajectories of autoignition simulations.

To standardize testing and the Pyrometheus API across languages, bindings from \Cpp{} and Fortran to Python are used to maintain a single backend-agnostic test suite written in Python (as opposed to writing the tests in all supported languages).
Standard build and interfacing tools generate the bindings and dynamic module imports in Python ensure that the tests interact with each backend.
As a result, the \Cpp{} and Fortran backends are as well tested as the Python code.
Each pull request to the GitHub repository tests all backends.

\begin{figure}[ht]
    \begin{center}
        \includegraphics[scale=1,page=1]{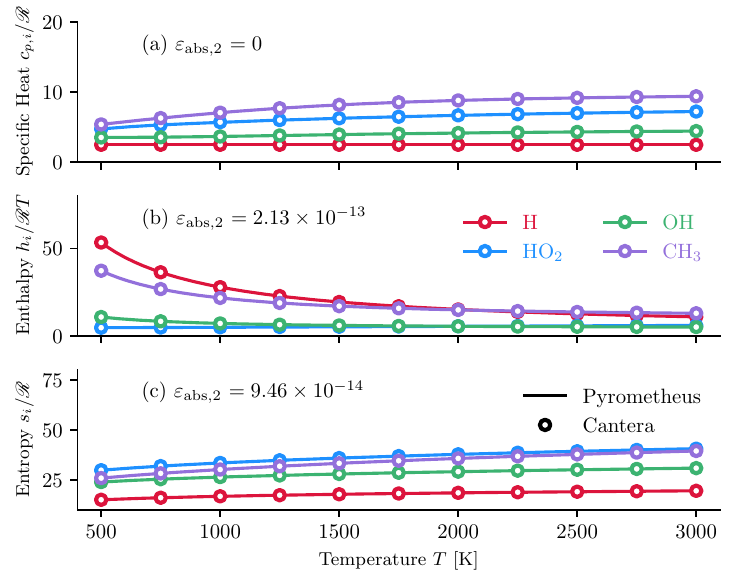}
        \caption{Thermodynamic properties for selected key radicals in the GRI-3.0 mechanism: (a) specific heat at constant pressure, (b) enthalpy, and (c) entropy.
        The reported error is averaged over all species in the mechanism and follows \cref{eq:verif_errors} with $p = 2$.}       
        \label{fig:pyro_verif_thermo}
    \end{center}
\end{figure}

\begin{figure}[ht]
    \begin{center}
        \includegraphics[scale=1,page=1]{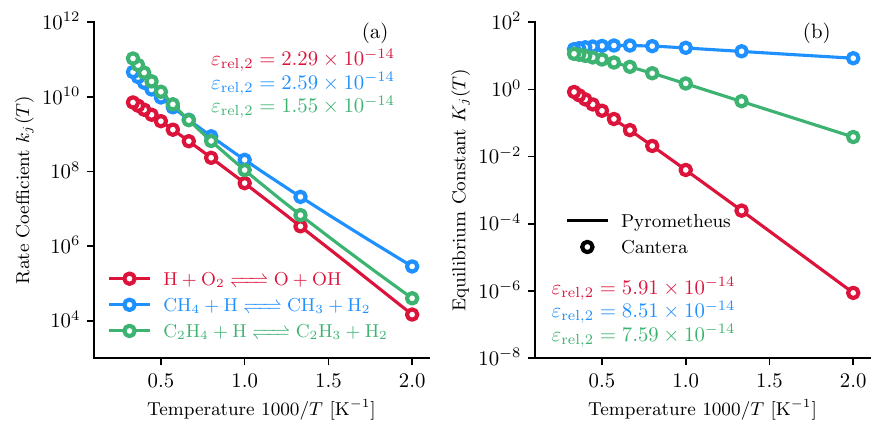}
        \caption{Reaction properties for selected key reactions in the GRI-3.0 mechanism: (a) reaction rate coefficient \cref{eq:rate_coeff}, and (b) equilibrium constant \cref{eq:equil_constants}.
        Reported errors are based on \cref{eq:verif_errors} with $p = 2$.}        
        \label{fig:pyro_verif_kinetics}
    \end{center}
\end{figure}

\begin{figure}[ht]
    \begin{center}
        \includegraphics[scale=1,page=1]{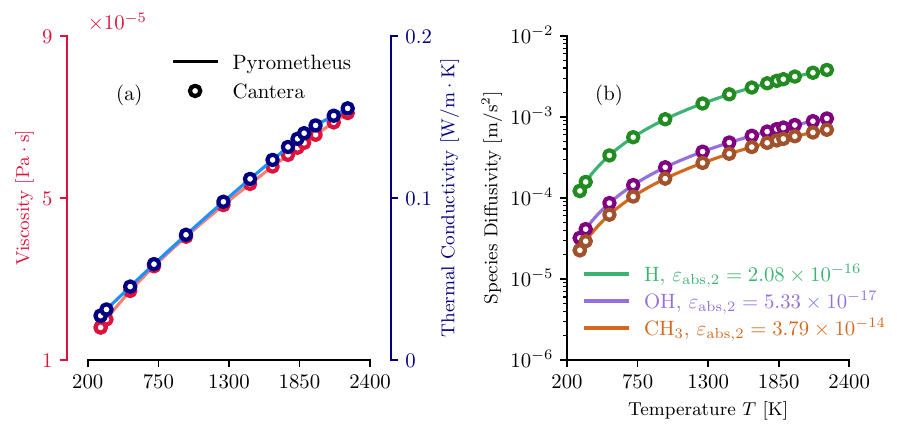}
        \caption{Mixture-averaged transport properties for a laminar methane--air free flame at $\SI{101325}{\pascal}$: (a) mixture viscosity and thermal conductivity; and (b) diffusivities of selected species in the GRI-3.0 mechanism.
        Reported errors follow \cref{eq:verif_errors} with $p = 2$.}
        \label{fig:pyro_verif_transport}
    \end{center}
\end{figure}

Errors are in the order of machine precision, as no approximations are made with respect to Cantera.
For arbitrary thermochemical quantity $\vec{\phi}$ (such as species production rates in \cref{eq:production_rates}), conventional absolute and relative errors
\begin{equation}\label{eq:verif_errors}
    \varepsilon_{\mathrm{abs},p}(\vec{\phi}) \equiv \| \vec{\phi}_{\mathrm{pyro}} - \vec{\phi}_{\mathrm{ct}} \|_{p}, \qquad 
    \varepsilon_{\mathrm{rel},p}(\vec{\phi}) \equiv \frac{ \| \vec{\phi}_{\mathrm{pyro}} - \vec{\phi}_{\mathrm{ct}} \|_{p} }{ \| \vec{\phi}_{\mathrm{ct}} \|_{p} },
\end{equation}
quantify code correctness, with $\vec{\phi}_{\mathrm{pyro}}$ and $\vec{\phi}_{\mathrm{ct}}$ the results from Pyrometheus and Cantera, and the vector dimension typically representing temperature.
Absolute errors in thermodynamic properties of selected species---as predicted by NASA polynomials \cref{eq:nasa_poly,eq:nasa_poly_c}---are shown in \cref{fig:pyro_verif_thermo}.
Rate coefficients and equilibrium constant, with corresponding errors \cref{eq:verif_errors}, are shown in \cref{fig:pyro_verif_kinetics} for selected reactions and species in the GRI-3.0 mechanism.
Correctness of mixture-averaged transport properties is demonstrated using states from a one-dimensional freely propagating flame stoichiometric methane--air flame at $\SI{101325}{\pascal}$ (computed with Cantera).
Viscosity \cref{eq:mixavg_viscosity}, thermal conductivity \cref{eq:mixavg_conductivity}, and the diffusivity \cref{eq:diff_mix_rule} of exemplar species are shown in \cref{fig:pyro_verif_transport}.

\section{Demonstration I: Computation via Deferred Addition of Details}\label{sec:results}

\subsection{Aim of Demonstrations}\label{subsec:aim_demo}

We demonstrate the use and advantages of the computational approach of \cref{sec:design} for combustion simulations, applied to the full formulation of \cref{sec:formulation}.
The demonstrations follow the workflow of \cref{fig:pyro_workflow}: from symbolic (on the leftmost edge) to the capabilities expressed on the rightmost edge by adding details only when needed.
We show how the same generated code can process numeric data using NumPy, be automatically differentiated with special-purpose array libraries such as JAX, and transformed for computation on GPU devices using the symbolic array library Pytato.
Homogeneous reactors and flames are simulated, and advanced applications to sensitivity analysis and computational cost modeling are discussed.


\subsection{Python Code Generation and Immediate Execution using NumPy}\label{subsec:numpy}

The initial step to computation is the generation of Python code from the symbolic representation.
The key input to code generation is a chemical mechanism and its parameters, encapsulated in the Cantera solution object of \cref{l:python_codegen}, line~1.
Here, we use the hydrogen--air San Diego mechanism~\cite{ref:Saxena2006}, as it is sufficient to demonstrate our approach and achieve our goals.
The Cantera solution interface populates mechanism-specific parameters in our symbolic representation.
We then map the representation to Python, as described in \cref{subsec:symbolic}.
This results in a computational representation of the thermochemistry that processes arrays, though it is not constrained to a specific array dimension, shape, or data type (numeric or symbolic).
The code is general in this sense, and the following examples showcase distinct capabilities derived from it without modification or re-generation.

\begin{lstlisting}[
float,
floatplacement=ht,
language=Python,
basicstyle={\footnotesize\ttfamily\color{myBlue}},
linerange={1-2},
label={l:python_codegen}, 
caption={Python thermochemistry code generation.}
]
sol = ct.Solution('sandiego.yaml')
pyro_cls = pyro.codegen.python.get_thermochem_class(sol)
\end{lstlisting}

In Python, Pyrometheus generates a class; computation requires that it be instantiated into an object.
As stated in \cref{subsec:ir_gpus}, the class constructor expects a NumPy-like library that provides the rules of computation for the array-based expressions in the generated code.
The established library NumPy provides these rules for numerical data, so it can be used to execute the methods of the generated code immediately.
A NumPy-based Pyrometheus object is instantiated in \cref{l:pyro_numpy_class} (the \lstinline|make_pyro_object| function that returns the object resolves minor syntactic differences in how arrays are concatenated across multiple array libraries).
We next demonstrate computation with this object, from illustrative calculations to simulations of reacting flows.

\begin{lstlisting}[
float,
floatplacement=ht,
language=Python,
basicstyle={\footnotesize\ttfamily\color{myBlue}},
linerange={4-5},
label={l:pyro_numpy_class}, 
caption={NumPy-based instance of the generated code for immediate calculation without transformations.}
]
import numpy as np
pyro_gas = make_pyro_object(pyro_cls, np)
\end{lstlisting}

An illustrative computation of the net production rates \cref{eq:production_rates} is shown in \cref{l:numpy_quick_computation}.
For this initial demonstration, the inputs are scalar density and temperature, as well as species mass fractions represented by simple one-dimensional arrays.
The more complex case of data on a grid for multi-dimensional combustion will be shown subsequently.
Line~8 shows the call to the specific method that computes~$\vec{\dot{\omega}}$ from these inputs, and line~12 compares the result against Cantera.
We stress that these code listings are part of a broader tutorial distributed with the code and can be run by users.

\begin{lstlisting}[
float,
floatplacement=ht,
language=Python,
basicstyle={\footnotesize\ttfamily\color{myBlue}},
linerange={7-18},
label={l:numpy_quick_computation}, 
caption={Example computation of the net production rates. Here, the density and temperature are scalars, and the mass fractions are one-dimensional arrays of length $N_{s} = 9$ for the hydrogen--air San Diego mechanism.}
]
temperature = 1200
sol.TP = temperature, pyro_gas.one_atm
sol.set_equivalence_ratio(phi=1, fuel='H2:1',
                          oxidizer='O2:0.21, N2:0.79')
mass_fractions = pyro_gas._pyro_make_array(sol.Y)

# Compute the density and molar net production rates
density = pyro_gas.get_density(pyro_gas.one_atm, temperature, mass_fractions)
omega = pyro_gas.get_net_production_rates(density, temperature, mass_fractions)

# Compare against Cantera
assert pyro_gas.usr_np.linalg.norm(omega - sol.net_production_rates) < 1e-14
\end{lstlisting}

The generated code also operates on multi-dimensional arrays without modification.
Two~dimensional inputs are shown in \cref{l:numpy_multidim_inputs}.
While obviously different to the inputs of \cref{l:numpy_quick_computation}, the interface to the computation of $\vec{\dot{\omega}}$ remains as lines~8--9 in \cref{l:numpy_quick_computation}.
This embodies the separation of concerns in the programming strategy: the burden of array details falls on the array library (NumPy in this case), not the generated code.
This cannot be done with thermochemistry libraries, which impose constraints on their input types and require nested calls within loops to compute across multiple points. 

\begin{lstlisting}[
float,
floatplacement=ht,
language=Python,
basicstyle={\footnotesize\ttfamily\color{myBlue}},
linerange={20-29},
label={l:numpy_multidim_inputs}, 
caption={Example computation of the net production rates for multi-dimensional arrays that represent thermochemistry data structured on a grid.}
]
num_x = 256
num_y = 256
mass_fractions = pyro_gas._pyro_make_array([
    pyro_gas.usr_np.tile(y, (num_y, num_x)) for y in sol.Y
])
temperature = 1200 * pyro_gas.usr_np.ones((num_y, num_x))

# Compute the density and molar net production rates
density = pyro_gas.get_density(pyro_gas.one_atm, temperature, mass_fractions)
omega = pyro_gas.get_net_production_rates(density, temperature, mass_fractions)
\end{lstlisting} 

\subsubsection{Application: Simulations of Autoignition and Flame Propagation}\label{subsub:app_sims}

\begin{figure}[ht]
  \begin{center}
    \includegraphics[scale=1,page=1]{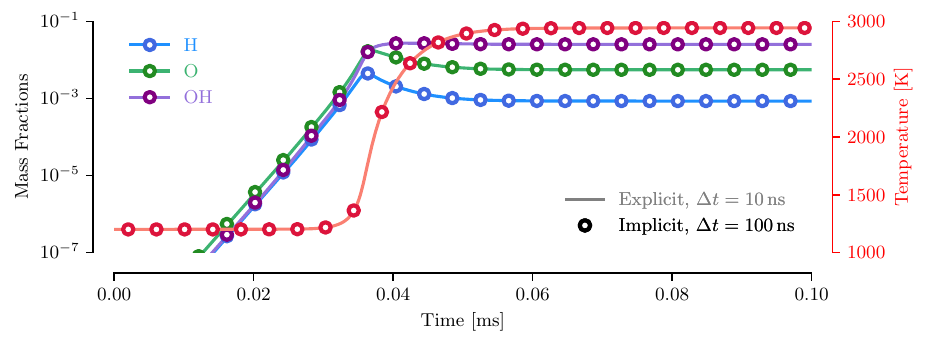}
    \caption{
    Thermochemical state during auto-ignition of a stoichiometric hydrogen--air mixture. Simulations using Pyrometheus-generated Python code.
    The explicit time marching uses NumPy-based Pyrometheus thermochemistry, while the implicit scheme uses JAX-based thermochemistry to enable AD for Jacobians.
    }
    \label{fig:pyro_autoign_results}
  \end{center}
\end{figure}

With the NumPy backend of Pyrometheus~(\cref{l:pyro_numpy_class}, line~2), we simulate autoignition in a stoichiometric hydrogen--air homogeneous reactor by evolving mass fractions in time per \cref{eq:massfrac_ode} via explicit time stepping with standard fourth-order accurate explicit Runge--Kutta.
Results are shown in \cref{fig:pyro_autoign_results}; a time step $\Delta t = \SI{10}{\nano\second}$ is used, with larger time steps lead to unstable simulations. This demonstrates that even without advanced AD capabilities, Pyrometheus can be used to simulate combustion using NumPy.   

We use the capability to process multi-dimensional arrays to simulate flames.
In particular, we simulate freely propagating laminar flames in one dimension.
For this, we numerically solve the compressible flow equations with chemical reactions~\cite{ref:DiRenzo2020}, using the schemes of \cref{subsec:time_integ}.
The system includes the conservation of species mass densities \cref{eq:massfrac_pde}, along with equations for the conservation of momentum density and total energy density. The $L = \SI{0.08}{\meter}$ domain is discretized using $2048$ points, and the $x = 0$ and $x = L$ boundaries are nonreflecting outflows via characteristic boundary conditions~\cite{ref:Poinsot1992}. The pressure is \SI{101325}{\Pa} throughout. The initial mixture is divided into two regions: a $T = \SI{1600}{\K}$ region of equilibrium products (at constant temperature and pressure), and a $T = \SI{300}{\K}$ region of unburned reactants in stoichiometric proportion. The two regions are smoothly connected via a hyperbolic tangent.

The resulting solver inherits the advantages of Pyrometheus, expressing just the equations without loops or other entangling computational details.
We show this property in \cref{l:pyro_eos_fn}, which
presents the routine to compute primitive from conserved variables. As seen, the routine expresses only the mathematical operations. NumPy handles array processing, so computational concerns have been separated.

\begin{lstlisting}[
float,
floatplacement=ht,
language=Python,
basicstyle={\footnotesize\ttfamily\color{myBlue}},
label={l:pyro_eos_fn}, 
caption={Routine to evaluate the equation of state in flame simulations. It expresses only the mathematical operations, without loops or other computational details. The type \lstinline|FlameState| in the function declaration is a container for NumPy arrays. A concrete implementation is provided with the release.}
]
def equation_of_state(self, cons_vars: FlameState) -> FlameState:
    """
    Evaluate primitive variables from conserved variables.
    """
    density = self.pyro_gas.usr_np.sum(cons_vars.densities, axis=0)
    mass_fractions = cons_vars.densities / density
    velocity = cons_vars.momentum / density
    energy = cons_vars.total_energy / density - 0.5 * velocity ** 2

    temperature = self.pyro_gas.get_temperature(energy, self.temp_guess,
                                                mass_fractions, do_energy=True)
    pressure = self.pyro_gas.get_pressure(density, temperature, mass_fractions)
    return FlameState(velocity, pressure, mass_fractions), density, temperature
\end{lstlisting}

\begin{figure}[ht]
  \begin{center}
    \includegraphics[scale=1,page=1]{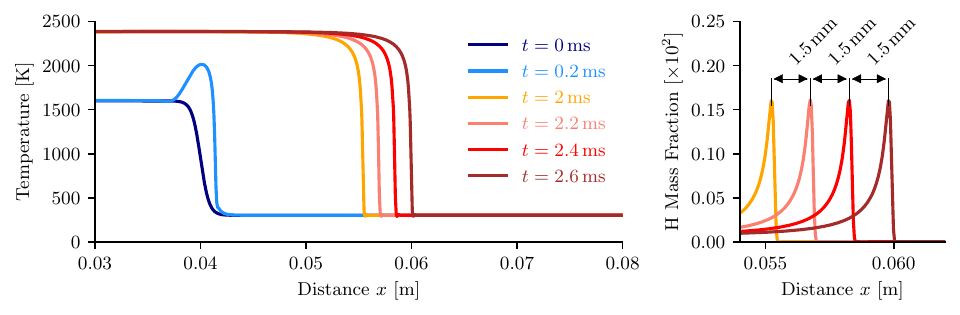}
    \caption{
    Propagation of an freely propagating laminar flame simulated using Pyrometheus-generated Python code: (a) temperature, and (b) \ce{H} mass fraction. The pressure is \SI{101325}{\pascal} and the equivalence ratio is $1$. After an initial transient, the flame ignites and propagates steadily into the unburned mixture. 
    }
    \label{fig:pyro_flame_results}
  \end{center}
\end{figure}

Flame simulation results are shown in \cref{fig:pyro_flame_results}.
After an initial transient, during which the mixture ignites, a flame is established and propagates into the fresh, cold mixture at a constant speed of $\SI{2.43}{\meter\per\second}$.
This is within $4.01\%$ of the flame speed predicted by Cantera's steady low-Mach number solver.
This error is smaller than the difference in flame speed across different transport models and can be attributed to small velocity and pressure fluctuations in the compressible flow solver. 

The examples we have presented demonstrate how to use the generated code instantly for multiple use cases.
Beyond their illustrative value, these examples can serve researchers working on novel chemical mechanisms by providing scripting access to kinetic properties such as ignition time and flame speed.
The following examples extend deeper into our pipeline to computation via symbolic array libraries.

\subsection{Automatic Differentiation Enables Implicit Time Marching}\label{subsec:pyro_ad}

The preceding demonstrations focused on the instant execution of the generated code achieved via pairing with NumPy.
We now focus on cases that require exact derivatives.
These are computed with automatic differentiation tools.
As stated in \cref{subsub:ad_arrays}, these tools provide extended meaning for array expressions. In practice, they construct the data-flow graph and propagate derivatives along its edges to compute gradients.

The starting point is the generated class of \cref{l:pyro_numpy_class}.
From it, we create instances that can be automatically differentiated.
Libraries developed for scientific machine learning, such as Google JAX and PyTorch, serve this purpose.
We just use JAX here, as shown in \cref{l:pyro_jaxnumpy_class}; PyTorch can be used in the same way.

\begin{lstlisting}[
float,
floatplacement=ht,
language=Python,
basicstyle={\footnotesize\ttfamily\color{myBlue}},
linerange={1-2},
label={l:pyro_jaxnumpy_class}, 
caption={JAX-based Pyrometheus instance for automatic differentiation.}
]
import jax.numpy as jnp
pyro_gas = make_pyro_object(pyro_cls, jnp)
\end{lstlisting}

To construct the data-flow graph, we must ensure data-independent control flow.
The only iterative procedure in thermochemistry computation is the inversion of temperature from internal energy and mass fractions.
In the generated code for the full thermochemistry formulation, this method is coded exactly as in \cref{subsec:temp_newton}, subject to termination by data-dependent tolerance.
When working with JAX, we re-define this method, as shown in \cref{l:pyro_jax_get_temp}, to ensure the that the DFG can be constructed.

\begin{lstlisting}[
float,
floatplacement=ht,
language=Python,
basicstyle={\footnotesize\ttfamily\color{myBlue}},
%linerange={61-77},
label={l:pyro_jax_get_temp}, 
caption={Data-independent temperature inversion, implemented using JAX constructs.}
]

\end{lstlisting}

\begin{figure}[ht]
    \begin{center}
        \includegraphics[scale=1,page=1]{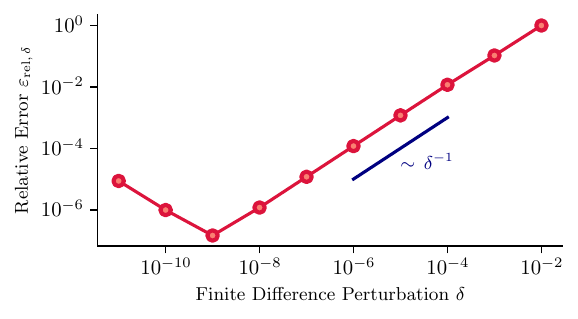}
        \caption{Relative error \cref{eq:jac_error} in the Jacobian \cref{eq:chem_jacobian} calculated using AD and first-order finite difference with varying perturbation size $\delta$. }
        \label{fig:pyro_ad_jacobian}
    \end{center}
\end{figure}

We proceed to compute derivatives automatically.
This computation is demonstrated in \cref{l:pyro_jax_ad} for the chemical source terms $\vec{S}$ in \cref{eq:massfrac_ode}.
To obtain their Jacobian
\begin{equation}\label{eq:jacobian}
    \mathbf{J} \equiv \frac{\partial \vec{S}}{\partial \vec{Y}},
\end{equation}
with $\vec{Y}$ the mass fractions, JAX builds the computational graph in line~12 of \cref{l:pyro_jax_ad} for the function defined in lines~4--9.
It then applies the chain rule as it traverses the graph in line~13.
The mixture state for which the Jacobian is computed is obtained from the explicit auto-ignition simulation of \cref{fig:pyro_autoign_results} at $t = \SI{0.0375}{\milli\second}$.
In \cref{fig:pyro_ad_jacobian}, the AD Jacobian $\mathbf{J}_{\mathrm{AD}}$ is compared against a numerical approximation $\mathbf{J}_{\mathrm{FD}}(\delta)$ based on first-order finite differences with varying perturbation size $\delta$ using 
\begin{equation}\label{eq:jac_error}
    \varepsilon_{\mathrm{rel},\delta} \equiv \frac{\| \mathbf{J}_{\mathrm{AD}} - \mathbf{J}_{\mathrm{FD}}(\delta) \|_{F}}{\| \mathbf{J}_{\mathrm{AD}} \|_{F}},
\end{equation}
where $\| \cdot \|_{F}$ is the Frobenius norm.
As expected, the error decreases linearly with finite-difference perturbation size until it is overcome by machine round-off errors for $\delta > 10^{-9}$.

\begin{lstlisting}[
float,
floatplacement=ht,
language=Python,
basicstyle={\footnotesize\ttfamily\color{myBlue}},
linerange={4-18},
label={l:pyro_jax_ad}, 
caption={
    Automatic differentiation of the chemical source terms using JAX. The temperature inversion routine is that of \cref{l:pyro_jax_get_temp}.
    The mixture state is obtained from the explicit autoignition simulation of \cref{fig:pyro_autoign_results} at $\SI{0.0375}{\milli\second}$.}
]
temp_guess = 1200
density, energy, mass_fractions = from_ignition_file(pyro_gas)

def chemical_source_term(mass_fractions):
    temperature = pyro_gas.get_temperature(energy, temp_guess, mass_fractions)
    return (
        pyro_gas.molecular_weights *
        pyro_gas.get_net_production_rates(density, temperature, mass_fractions)
    )

from jax import jacfwd
jacobian = jacfwd(chemical_source_term)
j_ad = jacobian(mass_fractions)
w_ad = chemical_source_term(mass_fractions)
\end{lstlisting}

Next, we use the JAX-based Pyrometheus instance to predict Hydrogen--air autoignition, as we did in \cref{subsub:app_sims} using NumPy, though now with the implicit scheme of \cref{subsec:time_integ}.
This scheme uses chemical Jacobians and allows a factor of 10 larger time step sizes than standard fourth-order accurate Runge--Kutta schemes.
\Cref{fig:pyro_autoign_results} shows that temperature and species profiles match those predicted with the explicit scheme. 

\subsection{Code Transformations Enable Execution on GPUs}\label{subsec:pyro_gpus}

Our next demonstration shows how, through code transformations, the Python class of \cref{l:python_codegen} can be used for execution on GPUs.
These devices leverage multiple logical units to perform the same calculation over large datasets efficiently.
We seek a representation for computation on GPUs, and languages like OpenCL and CUDA provide the necessary access to their low-level constructs.
The following approach transforms the Python code into these languages, achieved by specifying a suitable array library.
Importantly, we maintain access to every intermediate representation throughout the process.
This degree of control is unavailable with machine learning libraries like JAX and PyTorch, which can also be used to execute on GPUs.

We quantify performance via a roofline cost model~\cite{ref:Roofline} to relate performance to memory traffic. 
The roofline represents hardware-imposed performance bounds, and optimizations (such as parallelization) represent intermediate ceilings.
We quantify kernel performance via its nearness to the device roofline, whether memory- or compute-bound.

Recall that the Python class of \cref{l:python_codegen} is not tied to particular input data or hardware.
As depicted in \cref{fig:pyro_workflow}, it can still be thought of as a symbolic representation.
To retrieve the DFG from the generated Python code, we pair it with the symbolic Pytato library, as shown in \cref{l:pyro_pytato_class}.
Pytato arrays are shown in \cref{l:pyro_pytato_arrays}.
These arrays are related to \cref{l:lazy_arrays} of \cref{subsub:lazy_arrays}. The Pytato arrays are defined through an identifier, the shape, and the element data type; there are no numerical data behind them.
The symbolic arrays are inputs to the thermochemistry routines in \cref{l:pyro_pytato_graph}.
As these are executed, the state of the program is recorded into the corresponding DFG. 

\begin{lstlisting}[
float,
floatplacement=ht,
language=Python,
basicstyle={\footnotesize\ttfamily\color{myBlue}},
linerange={1-2},
label={l:pyro_pytato_class}, 
caption={Pytato-based Pyrometheus instance to retrieve the data flow graph from the generated Python code. Pytato provides lazy evaluation of array-based expressions, as discussed in \cref{subsub:lazy_arrays}.}
]
import pytato as pt
pyro_gas = make_pyro_object(pyro_cls, pt)
\end{lstlisting}

Our goal is to evaluate chemical source terms $\{ S_{i} \}_{i = 1}^{N}$, as defined in \cref{eq:massfrac_ode}, using accelerators.
As stated in \cref{subsec:ir_gpus}, the computational graph is tied to the inputs to the computation.
Depending on the application, there are two possible sets of inputs.
In the isothermal case, the inputs are mass fractions and temperature.
This case is relevant to flame simulations where, as shown in \cref{l:pyro_eos_fn}, the temperature is obtained as part of the EOS evaluation.
In the adiabatic case, the inputs are the mass fractions and internal energy (or enthalpy)---from which the temperature is inverted, as shown in \cref{l:pyro_jax_ad}.
This case is relevant to auto-ignition simulations at constant energy and density (or enthalpy and pressure).
We focus on the isothermal case due to its importance for large-scale flame simulations.
The adiabatic case is smaller in size and requires an extension to handle data-dependent control flow, which is not implemented in Pytato and discussed further in \cref{sec:impact}. 


\begin{lstlisting}[
float,
floatplacement=ht,
language=Python,
basicstyle={\footnotesize\ttfamily\color{myBlue}},
linerange={4-9},
label={l:pyro_pytato_arrays}, 
caption={Symbolic input arrays to construct the data flow graph for the net production rates (shown in \cref{l:pyro_pytato_graph}).}
]
inner_length = 32
num_x = 32 * inner_length
num_y = 32 * inner_length
density = pt.make_placeholder(name='density', shape=(num_x, num_y), dtype='float64')
temperature = pt.make_placeholder(name='temperature', shape=(num_x, num_y), dtype='float64')
mass_fractions = pt.make_placeholder(name='mass_fractions', shape=(pyro_gas.num_species, num_x, num_y), dtype='float64')
\end{lstlisting}

The Pyrometheus-based routine to evaluate the chemical source terms is shown in \cref{l:pyro_pytato_graph}.
While NumPy-like, this method is unconstrained to specific input types.
For the symbolic inputs of \cref{l:pyro_pytato_arrays}, the routine returns a Pytato array expression akin to those in \cref{l:lazy_arrays}, from which the computational graph is assembled as a Python dictionary (lines~8--10).
This graph expresses arithmetic relations between arrays and is to be transformed to incorporate loops and parallelism.
However, it can be used beyond this purpose.
For example, it can be used to create computational cost models by counting floating point operations. 

\begin{lstlisting}[
float,
floatplacement=ht,
language=Python,
basicstyle={\footnotesize\ttfamily\color{myBlue}},
linerange={11-20},
label={l:pyro_pytato_graph}, 
caption={Construction of data flow graph \lstinline|pyro_graph| using the symbolic arrays of \cref{l:pyro_pytato_arrays}.}
]
def chemical_source_term(density, temperature, mass_fractions):
    omega = pyro_gas.get_net_production_rates(density, temperature, mass_fractions)
    return pyro_gas._pyro_make_array([
        w * om for w, om in zip(pyro_gas.molecular_weights, omega)
    ])

omega = chemical_source_term(density, temperature, mass_fractions)
pyro_graph = pt.make_dict_of_named_arrays({
    f'omega{i}': w for i, w in enumerate(omega)
})
\end{lstlisting}

In \cref{l:pyro_loopy}, the computational graph of \cref{l:pyro_pytato_graph} is translated to a Loopy kernel (\cref{subsec:ir_gpus}).
The Loopy kernel extends the arithmetic expressed by the computational graph with data access patterns represented by loop variables. At this stage, any code generated from the kernel in line~1 of \cref{l:pyro_loopy} will execute the instructions within its loops sequentially.
To target a more efficient code, we imbue parallelization into the kernel.
Following the OpenCL--CUDA model of computation, the nested loop is partitioned into outer and inner groups per \cref{eq:lp_split}, as shown in lines 6--13 of \cref{l:pyro_loopy}.
Outer groups represent global work groups (or blocks in CUDA), while inner groups represent work items (or threads in CUDA).
With the Loopy commands of \cref{l:pyro_loopy} lines~15--16, CUDA code is generated, which can be executed via the PyCUDA packages.

\begin{lstlisting}[
float,
floatplacement=ht,
language=Python,
basicstyle={\footnotesize\ttfamily\color{myBlue}},
linerange={1-16},
label={l:pyro_loopy}, 
caption={
    The data flow graph for the chemical source terms $\{S_{i}\}_{i = 1}^{N}$ is mapped onto a Loopy kernel.
    The loops are split into global and local domains of dimension $(32\times 32)$.
    In CUDA, this dimension corresponds to the number of threads in each direction; in OpenCL, it corresponds to the number of local work items per global work group.
}
]
pyro_kernel = pt.generate_loopy(
    pyro_graph,
    function_name='chemical_source_terms',
).program
...
pyro_kernel = lp.split_iname(
    pyro_kernel, 'i', inner_length,
    outer_tag='g.0', inner_tag='l.0'
)
pyro_kernel = lp.split_iname(
    pyro_kernel_split, 'j', inner_length,
    outer_tag='g.1', inner_tag='l.1'
)

pyro_kernel = pyro_kernel.copy(target=lp.CudaTarget())
code_str = lp.generate_code_v2(pyro_kernel).device_code()
\end{lstlisting}

\begin{figure}[ht]
  \begin{center}
    \includegraphics[scale=1,page=1]{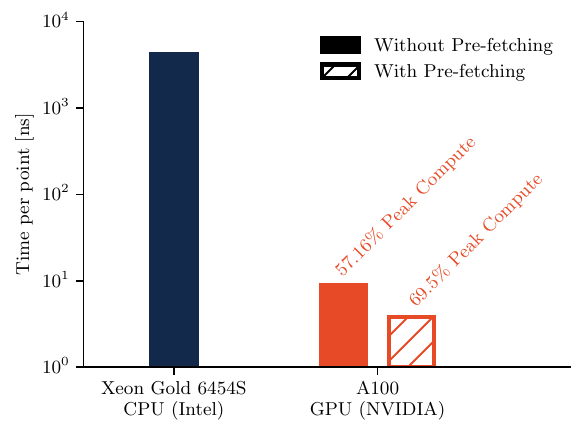}
    \caption{
    Compute time per point for hydrogen--air chemical source terms $\{ S_{i} \}_{i = 1}^{N}$ given temperature and mass fractions.
    Smaller is faster. 
    The problem size is $(1024,\, 1024)$ points.
    Each device executes $(32,\, 32)$ threads per block.
    The data prefetching transformation halves execution time on the NVIDIA A100 GPU, reaching $69.5\%$ of the device's peak fused multiply--add double precision performance.
    }
    \label{fig:pyro_gpu_performance}
  \end{center}
\end{figure}

Computational performance is shown in \cref{fig:pyro_gpu_performance} for the Intel~Xeon Gold CPU (6454S) and the NVIDIA~A100 GPU.
The CPU case was run with OpenCL-generated code compiled via PoCL, while the GPU case was run with CUDA-generated code and compiled with the nvhpc~24.5 toolkit.
The problem size is $(1024\times1024)$ points, with $(32\times32)$ work groups (blocks) and $(32\times32)$ work items (threads).
The GPU achieves a time per point of $\SI{8.86}{\nano\second}$, a factor of $425.81$ faster than the CPU.
This performance is $57.1\%$ peak compute rate.

Additional transformations can be applied to improve GPU performance.
Data access patterns can be prefetched.
Data are moved from global to local memory and registers in each logical unit before they are needed, mitigating latency and improving the rate at which operations are conducted.
This memory management technique can be expressed in both OpenCL and CUDA, and Loopy provides a corresponding transformation, shown in \cref{l:pyro_loopy_fetch}.
Data prefetching on CPU devices has no apparent effect on time per point, as this optimization is likely exposed to the compiler.
However, on the A100~GPU, prefetching improves performance by a factor of $2.3$, or $69.5\%$ of peak compute on the device.  

\begin{lstlisting}[
float,
floatplacement=ht,
language=Python,
basicstyle={\footnotesize\ttfamily\color{myBlue}},
linerange={18-21},
label={l:pyro_loopy_fetch}, 
caption={The parallelized kernel of line 15 in \cref{l:pyro_loopy} is further transformed to prefetch the input data according to its access patterns within the loops. During execution, the program moves data from global to local memory before it is needed, improving the compute rate.}
]
pyro_kernel = pyro_kernel
pyro_kernel = lp.add_prefetch(pyro_kernel, 'density',)
pyro_kernel = lp.add_prefetch(pyro_kernel, 'temperature',)
pyro_kernel = lp.add_prefetch(pyro_kernel, 'mass_fractions')
\end{lstlisting}

The developed compilation pipeline is open to the user from symbolic to OpenCL and CUDA.
Corresponding PyOpenCL and PyCUDA packages handle further compilation while letting the user maintain control through Python.
We have demonstrated the advantages of providing access to additional transformations, such as data reuse.
Following \cref{subsub:lazy_arrays}, the Pytato--Loopy and subsequent transformations can be abstracted to a high-level interface that compiles the generated Python code.
This way, users can interact with high-performance code through a high-level control layer with a low barrier to entry.

The generated Python code is general.
The code can simulate the auto-ignition of flames (with explicit and implicit schemes) by separating algebraic and computational concerns.
It also transfers the burden of defining array-based semantics to an array library.
The code is compatible with numeric and symbolic array libraries such as NumPy and Pytato.
The generated code is portable, computing chemical source terms and thermodynamics on CPUs and GPUs.

\section{Demonstration II: Computation via Integration into Existing Solvers}\label{sec:chemfc}

The preceding demonstrations show a pipeline from symbolic representation to computation, using multiple code transformations and deferred detail addition.
Some applications do not benefit from retaining symbolic capabilities, requiring code that instead operates on numeric data immediately within a different control environment than Python.
This situation is common when using existing flow solvers.
Here, we demonstrate that Pyrometheus can generate code that is portable to existing solvers from the symbolic representation without executing the entire pipeline of \cref{sec:results}.

\subsection{Description of the MFC Flow Solver}

MFC is an open-source, MIT-licensed Fortran08 codebase.
MFC has been developed for the last two decades to advance understanding of multiphase flows (bubbles and droplets) and uses interface- and shock-capturing techniques~\citep{Bryngelson2020,wilfong252}.
A predecessor of MFC was the first to use the interface-capturing numerical approach to solve the five-equation model~\cite{Kapila2001}, which is a single velocity and energy equation reduced model of the multiphase model of \citet{Baer1986}.

For the past decade, MFC has been used to conduct high-fidelity multiphase flow simulations at extreme computational scales~\citep{bryngelson19_cpc,wilfong252}, including those of high density ratios~\citep{charalampopoulos21,bryngelson19_whales}.
Yet, MFC has not supported simulations of reacting flows due to the complexity of integration with exascale, performance-critical application development.
Here, we use our computational approach to create a thermochemistry representation for MFC that is compatible with its parallelization strategy.

The MFC algorithm stores and uses coalesced memory arrays for stencil reconstructions.
This strategy leads to the relatively high arithmetic intensity and performant code on GPU accelerators.
It employs a directional splitting approach that evaluates these quantities once per time step, then reuses them~\citep{rossinelli13,schmidmayer20,bryngelson19_cpc}.
MFC uses established computational methods, including high-order accurate interface reconstruction, approximate Riemann solvers, and Runge--Kutta time steppers.
MFC handles parallel processing via MPI3 and domain decomposition.

GPU offloading is accomplished via OpenACC.
MFC is performant on NVIDIA and AMD GPUs, including OLCF Summit (NVIDIA~V100) and OLCF Frontier and LLNL~El~Capitan machines (AMD~MI250X and MI300A).
MFC's results on such machines and its GPU offload implementation are documented elsewhere~\citep{radhakrishnan24,radhakrishnan22,wilfong24,elwasif23,wilfong252}.
Here, we demonstrate how the generated Fortran code from Pyrometheus is readily integrated into MFC.

\begin{figure}
    \centering
    \includegraphics[]{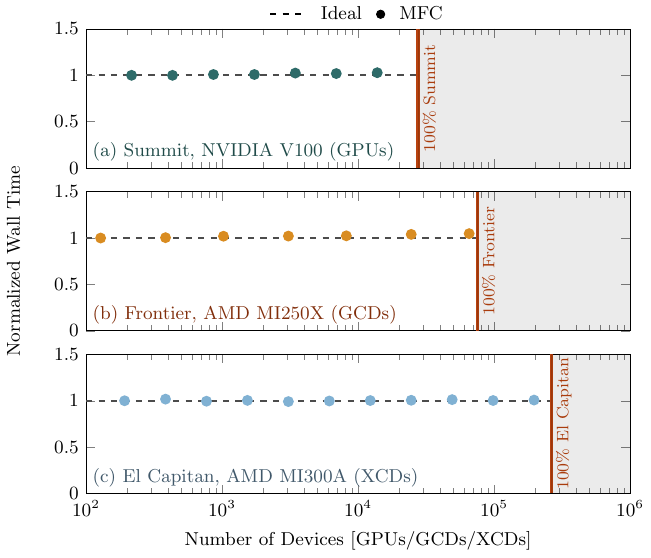}
    \caption{
        Weak scaling of MFC on LLNL~El~Capitan (AMD~MI300A XCDs) and OLCF~Frontier (AMD~MI250X GCDs) and Summit (NVIDIA~V100 GPUs) for a representative 3D structured-grid two-phase test problem.
    }
    \label{fig:chemfc-scaling}
\end{figure}

\Cref{fig:chemfc-scaling} shows how MFC scales to better than 60\% of all NVIDIA GPUs on OLCF~Summit, 88\% of the AMD~MI250X GPUs (GCDs) ($>66$K) on OLCF~Frontier (the first exascale computer), and 85\% of the AMD~MI300A APUs (XCDs) on LLNL~El~Capitan.
Each GPU computation is approximately 500~times faster than one on a comparable CPU device.

\subsection{Code Generation}

Thermochemistry code for MFC is generated directly from the symbolic representation of \cref{subsec:symbolic}.
Expressions are mapped to Fortran, and a template is rendered.
The interface to this process is shown in \cref{l:pyro_fortran_gen}, line~1, where the Cantera solution object is that of \cref{l:python_codegen}, line~1. 
The generated code is at the same abstraction level as MFC and will, therefore, form part of the same compilation unit as the flow solver.
As such, it is compatible with directive-based compile-time optimization.
The Pyrometheus Fortran template includes OpenACC directives, so it was readily compliant with MFC's (and many GPU-based codes) offloading strategies.

\begin{lstlisting}[
float,
floatplacement=t,
language=Python,
basicstyle={\footnotesize\ttfamily\color{myBlue}},
linerange={1-3},
label={l:pyro_fortran_gen}, 
caption={Fortran code generation for combustion thermochemistry. The resulting code is saved to a file to be compiled along with MFC.}
]
pyro_code = pyro.codegen.fortran90.gen_thermochem_code(sol)
with open(f'thermochem_{sol.name}.f90', 'w') as outf:
    print(pyro_code, file=outf)
\end{lstlisting}

\subsection{Performance of Pyrometheus-enabled MFC Simulations of Reacting Flows}

\begin{figure}
    \centering
    \includegraphics[]{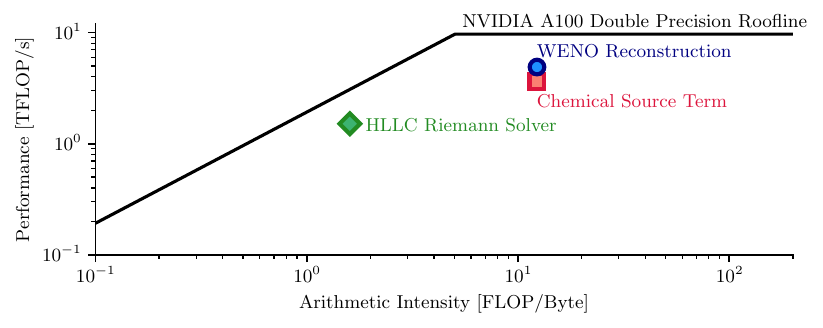}
    \caption{MFC kernel performance on an NVIDIA A100 GPU.
        The roofline corresponds to peak double precision fused multiply--add instructions.
        The kernel for chemical source terms computes $S_{i}$ in \cref{eq:massfrac_pde} for a given density, mass fractions, and internal energy.}
    \label{fig:mfc_roofline}
\end{figure}

Code performance is quantified using the roofline approach~\cite{ref:Roofline}.
We compare the cost of evaluating chemical source terms \cref{eq:production_rates} relative to that of kernels associated with fluxes, such as WENO and the HLLC approximate Riemann solver.
Because performance is independent of physical configuration, it is measured in three-dimensional simulations of auto-ignition in a quiescent hydrogen--air mixture, modeled using the hydrogen subset of the GRI-3.0 mechanism (with $10$~species, including Argon).
These are straightforward to configure, and results translate directly to more complex configurations, such as the detonation of \cref{subsec:detonation}.
Results, shown in \cref{fig:mfc_roofline}, are presented for a grid size of $17.98\times10^{6}$ cells, which corresponds to $84.5\%$ GPU memory utilization. Performance was averaged over kernel calls (with a total of $15$ per kernel).  

As seen in \cref{fig:mfc_roofline}, the computation of chemical source terms is close, by $6.04$~TFLOP/s, to the double-precision peak-FLOP ceiling.
This performance is comparable to the WENO kernel, which is $4.76$~TFLOP/s from the ceiling.
In comparison, the computation of chemical source terms outperforms the HLLC approximate Riemann solver, which, while still close to its optimal performance, is memory bound with an arithmetic intensity of $1.5$~FLOP/Byte.
Note that the WENO kernel is the most expensive computation in MFC: its time per cell is $\SI{137}{\milli\second}$ per call, a factor of $8.43$ larger than that of the source terms.

\subsection{Pyrometheus-enabled MFC Simulations of Detonations}\label{subsec:detonation}

\begin{figure}[H]
    \centering
    \includegraphics[scale=1]{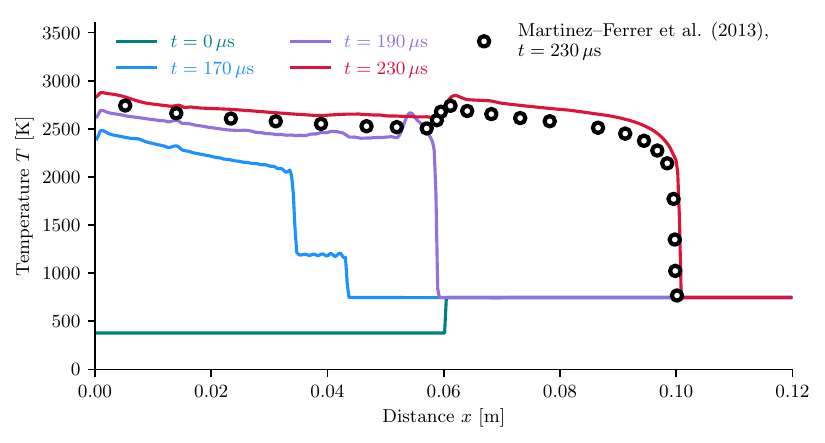}
    \caption{
        Temperature of a one-dimensional detonation in a \ce{H2}--\ce{Ar} mixture, simulated using MFC with Pyrometheus-generated Fortran code.
    }
    \label{fig:chemfc_detonation}
\end{figure}

To demonstrate the integration of Pyrometheus into MFC, we simulate a one-dimensional detonation propagating into a quiescent stoichiometric hydrogen--argon mixture.
This configuration is chosen because it couples flow and chemistry without complex boundary conditions and is well documented elsewhere~\cite{ref:Fedkiw1997,ref:Martinez2014}. 
Reactions are modeled using the hydrogen subset of the GRI-3.0 mechanism. The domain of length $L = \SI{0.12}{\meter}$ is discretized using $400$ cells.
The boundary at $x = 0$ is a reflecting wall, and the $x = L$ boundary is a non-reflecting outflow.
The flow is initialized with a shock at the center of the domain.
The shock propagates towards the wall, where it reflects and ignites the mixture as it propagates into the domain.
As seen in \cref{fig:chemfc_detonation}, a detonation forms after $\approx \SI{190}{\micro\second}$.
Our simulations reproduce the location of the detonation front at $t = \SI{230}{\micro\second}$.
The difference in temperature downstream of the wave with respect to the solution of \citet{ref:Martinez2014} is $\lesssim 5.7\%$.
This difference is small and can be attributed to the use of different chemical mechanisms by \citet{ref:Martinez2014}.

\subsection{Pyrometheus-enabled MFC Simulations of Flame--Vortex Interactions}

To demonstrate our capability for simulating thermochemistry and transport, we simulate a two-dimensional premixed hydrogen--air flame perturbed by two superimposed vortices.
This simulation closely follows the configuration of \citet{ref:Yu2012} and \citet{ref:Bando2020}.
In contrast with the previous one-dimensional detonation, this case incorporates mixture-averaged mass diffusion fluxes and corresponding transport coefficients \cref{eq:diff_mix_rule,eq:mixavg_viscosity,eq:mixavg_conductivity}.
Reactions are modeled using the hydrogen subset of the GRI-3.0 mechanism.
Fifth-order accurate WENO reconstruction and an HLLC approximate Riemann solver are used to compute the fluxes, following~\citep{wilfong252}.

\begin{figure}[h]
    \centering
    \includegraphics[scale=1.0]{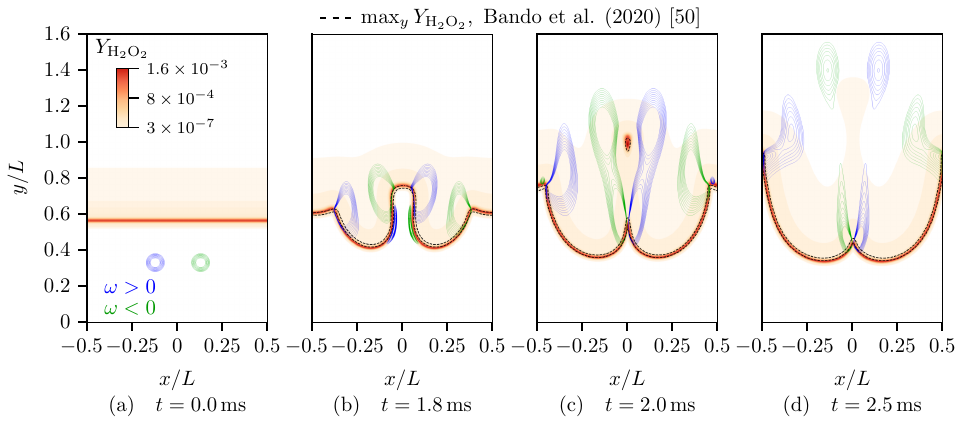}
    \caption{
        Flow configuration showing the evolution of flame--vortex interaction, represented by $\ce{H2O2}$ mass fraction and vorticity (blue and green contours, for positive and negative values) at (a) $t = \SI{1.8}{\milli\second}$, (b) $t = \SI{2.0}{\milli\second}$, and  (c) $t = \SI{2.5}{\milli\second}$.
        The dashed line denotes the flame front from \citet{ref:Bando2020}.        
    }
    \label{fig:chemfc_2D_YH2O2}
\end{figure}

\Cref{fig:chemfc_2D_YH2O2} shows the flow configuration.
The computational domain is $[0,\SI{8} {\milli\meter}]\times [0,\SI{16}{\milli\meter}]$, discretized with $(N_x, N_y) = (512, 1026)$ grid cells. 
Boundary conditions in $x$ are periodic; inflow and outflow boundary conditions are imposed at $y = 0$ and $y = 2L$.
The simulation is initiated with an extruded one-dimensional \ce{H2}--air flame with equivalence ratio $\phi=0.6$.
The flame front is positioned $\SI{4.7}{\milli\meter}$ from the right of the inflow boundary.
The vortices are superimposed $\SI{2.7}{\milli\meter}$ from the inflow, symmetrically arranged about $x = 0$ with $\SI{2}{\milli\meter}$ between their centers.

The flame--vortex interaction is shown in \cref{fig:chemfc_2D_YH2O2}.
We compare the flame shape with the results of \citet{ref:Bando2020}.
Close agreement in flame shape is observed.
Differences in absolute spatial position are due to the lack of sufficient geometric information in the reference solution.
Still, the vortex structures observed in our simulation also show good agreement with the results of \citet{ref:Yu2012}, where similar features in the vorticity field are reported.
Vortex structures downstream of the flame arise due to flame curvature, which induces baroclinicity.
The agreement in flame and flow structures supports the correctness of the implementation.

As stated, these simulations necessitate the computation of transport coefficients, which are provided to MFC by Pyrometheus.
This verification is in contrast with the simulation of one-dimensional detonation of \cref{fig:chemfc_detonation}.
For this hydrogen--air mechanism, the cost of calculating mixture-averaged transport properties is $\approx 4.2\%$ of total run time, which is $\approx 25\%$ of the time that it takes to assemble the corresponding mass diffusion fluxes.
This cost balance indicates the efficiency of our Pyrometheus implementation for transport.  

The simulations presented in this section showcase how Pyrometheus overcomes the challenges associated with coupling exascale flow solvers with computational models for chemical kinetics, thermodynamics, and transport.
In contrast with pre-compiled libraries, Pyrometheus places the computation of the thermochemistry and transport at the same compilation level as the flow solver.
MFC then offloads the fully coupled flow--combustion computation to GPUs using compiler directives (at odds with pre-compiled libraries), achieving near roofline performance (\cref{fig:mfc_roofline}). 

\section{Summary of Principal Features and Scope, with Additional Discussion}\label{sec:impact}

A computational approach for combustion thermochemistry has been developed.
The strategy is based on a symbolic representation rooted in syntactic rules, with the only assumption that the computation will eventually be array-based.
Together, the approach is sufficiently general to achieve multiple conflicting goals that are otherwise best served by disparate data structures.
Computational details, like loop patterns and parallelism, are added progressively.
With each addition, the representation is lowered closer to a case-specific target through code generation, rigorously presented via inference rules.

Two main compilation pipelines are presented.
The first preserves a control environment in Python throughout the calculation, and the second saves the code to a file.
This setup offers avenues for both modern and traditional high-performance computing.
It allows Python to continue expanding into HPC, notably with scientific machine learning, while supporting preexisting HPC solvers.

The achieved capabilities were demonstrated.
The generated code can simulate, for example, homogeneous reactors or flames.
It is compatible with AD libraries, which enable implicit time marching and sensitivity analysis. 
Its computation can be offloaded to GPUs via a diverse range of mechanisms.
Every compilation unit is exposed to the user, allowing the code to be analyzed and manipulated to improve performance.
This property also enables integration into CFD codebases such as MFC, as the generated code can be bundled with the flow solver for compilation and directive-based loop optimization.
This lower-level strategy contrasts against pre-compiled thermochemistry libraries like Cantera, which only expose parameters through an interface.

The computation of the thermochemistry involves data-dependent control flow in the inversion of the temperature from the internal energy.
Maintaining the control flow is required for an acyclic computational graph.
Pyrometheus addresses the control flow logistics via data-dependent loop abstractions (e.g., in JAX) or by limiting the number of Newton iterations.


The Pyrometheus approach can impact any large-scale combustion simulation because of the properties and capabilities of the generated code.
Application areas include turbulent combustion simulations, dynamic adaptive chemistry, where multiple sub-mechanisms need to be evaluated on the fly, machine learning of chemical source terms, model reduction, and uncertainty quantification, with additional array axes representing parametric realizations.
An early version of this approach has already been used for flow in a flameholder~\cite{ref:Cisneros2019}.

Beyond MFC, our implementation serves as the thermochemistry engine for multiple compressible-flow solvers.
The MIRGE-Com solver~\cite{ref:Ricciardi2024} is rooted in the same concepts and implements discontinuous Galerkin discretizations through symbolic manipulation to maintain the separation of concerns.
The PyFlowCL package~\cite{ref:Liu2024} implements data-driven sub-grid-scale models with PyTorch, enabled by Pyrometheus differentiable generated code.
The fully-differentiable JAX-Fluids solver~\cite{ref:Bezgin2023,ref:Bezgin2025} provides Pyrometheus as an alternative to its native implementation of thermochemistry.
PlasCom2~\cite{ref:Mikida2019,ref:Murthy2022,ref:Vollmer2022}, a \Cpp{} plasma-coupled combustion solver, uses an early version of our implementation.


\section{Conflict of Interest}

The authors have no known conflicts of interest associated with this publication, and there has been no significant financial support for this work that could have influenced its outcome.

\section*{Acknowledgments}

This material is based in part upon work supported by the Department of Energy, National Nuclear Security Administration, under Award Numbers DE-NA0002374 and DE-NA0003963.
The authors gratefully acknowledge Cory Mikida, Michael T.~Campbell, and Dr.~Tulio~Ricciardi for assisting in the implementation. ECG gratefully acknowledges Prof.\ Andreas~Kl\"{o}ckner, Dr.~Matthias Diener, and Matthew J.~Smith (University of Illinois) for extensive discussion on code generation, array semantics, and GPU computing.

SHB gratefully acknowledges support from the U.S. Department of Defense, Office of Naval Research under grant number N00014-24-1-2094.
Some computations were performed on the Tioga, Tuolumne, and El~Capitan clusters at Lawrence Livermore National Laboratory's Livermore Computing facility.
This research also used resources of the Oak Ridge Leadership Computing Facility---Frontier, Summit, and Wombat---at the Oak Ridge National Laboratory, which is supported by the Office of Science of the U.S. Department of Energy under Contract No.~DE-AC05-00OR22725 (allocation CFD154, PI~Bryngelson).
In addition, this work used Delta at the National Center for Supercomputing Applications and Bridges2 at the Pittsburgh Supercomputing Center through allocations PHY210084 and  PHY240200 (PI~Bryngelson) from the Advanced Cyberinfrastructure Coordination Ecosystem: Services \& Support (ACCESS) program, which is supported by National Science Foundation grants \#2138259, \#2138286, \#2138307, \#2137603, and \#2138296.

\appendix
\section{Combustion Thermochemistry Formulation \& Methods}\label{sec:formulation}

\subsection{Governing Equations}\label{subsec:gov_eqns}

A homogeneous adiabatic isochoric reactor is sufficient for introducing thermochemistry.
The chemical composition is characterized by the mass fractions $\{ Y_i \}_{i = 1}^{N}$ of $N$ species, which are governed by
\begin{align}\label{eq:massfrac_ode}
  &\rho\frac{dY_{k}}{dt} =
  \overbrace{ W_{k}\dot{\omega}_{k} }^{\equiv S_{k}},\qquad Y_{k}(0) = Y_{k}^{0},\qquad k = 1,\,\dots,\,N,
\end{align}
where $W_{k}$ is the molecular weight of the $k^{\mathrm{th}}$ species, $\dot{\omega}_{k} = \dot{\omega}_{k}(Y_{1},\, \dots,\, Y_{N},\, T)$ its net production rate (with $T$ the temperature), $\rho$ the density, and $\{ Y_{k}^{0} \}_{k = 1}^{N}$ the initial condition.
The right-hand side $S_{k}$ is typically referred to as the chemical source terms.
The pressure $p$ is obtained from the equation of state
\begin{equation}\label{eq:eos}
  p = \rho\,\mathscr{R}T / W,
\end{equation}
where
\begin{equation}
  W = {\bigg(} \sum_{i = 1}^{N}Y_{i}/W_{i} {\bigg)}^{-1}
\end{equation}
is the mixture molecular weight and $\mathscr{R}$ the universal gas
constant.
The focus of this work is the implementation and evaluation of $\dot{\omega}_{k}$, so corresponding detailed expressions are provided next.
Generalization to include advection and diffusion of the chemical species is briefly discussed in \cref{subsec:transport}.

Net production rates $\{ \dot{\omega}_{k} \}_{k = 1}^{N}$ represent changes in composition due to chemical reactions, which can be expressed as
\begin{equation}\label{eq:reactions}
  \sum_{i = 1}^{N}\nu_{i j}^{\prime}\mathcal{S}_{i} \rightleftharpoons
  \sum_{i = 1}^{N}\nu_{i j}^{\prime\prime}\mathcal{S}_{i},\qquad j =
  1,\dots,M,
\end{equation}
where $\nu_{ij}^{\prime}$ and $\nu_{ij}^{\prime\prime}$ are the forward and reverse stoichiometric coefficients of species $\mathcal{S}_{i}$ in the $j^{\mathrm{th}}$ reaction.
Per \cref{eq:reactions}, species $\mathcal{S}_{i}$ can only be destroyed or produced in proportion to $\nu_{ij}^{\prime}$ or $\nu_{ij}^{\prime\prime}$ in the $j^{\mathrm{th}}$ reaction.
Thus, $\{ \dot{\omega}_{i} \}_{i = 1}^{N}$ are linear combinations of the reaction rates of progress $R_{j}$,
\begin{equation}\label{eq:production_rates}
  \dot{\omega}_{i} = \sum_{j = 1}^{M}\nu_{ij}R_{j},\qquad i =
  1,\dots,N,
\end{equation}
where
\begin{equation}
  \nu_{ij} = \nu_{ij}^{\prime\prime} - \nu_{ij}^{\prime}
\end{equation}
is the net stoichiometric coefficient of the $i^{\mathrm{th}}$ species in the $j^{\mathrm{th}}$ reaction.
The rates of progress are given by the law of mass-action,
\begin{equation}\label{eq:reaction_rates}
  R_{j} = k_{j}(T)\psq{ 
  \prod_{\ell = 1}^{N}C_{\ell}^{\nu_{\ell j}^{\prime}} - 
  \frac{1}{K_{j}(T)}
  \prod_{k = 1}^{N}C_{k}^{\nu_{kj}^{\prime\prime}} 
  },\qquad j = 1,\dots,M,
\end{equation}
where
\begin{equation}\label{eq:concentrations}
  C_{k} = \rho Y_{k} / W_{k},\qquad k = 1,\dots,N,
\end{equation}
are the species molar concentrations, $k_{j}(T)$ is the rate coefficient of the $j^{\mathrm{th}}$ reaction and $K_{j}(T)$ its equilibrium constant.
Depending on the reaction, the rate coefficient $k_{j}(T)$ may take different forms (and even become a function of pressure).
Its simplest form is the Arrhenius expression,
\begin{equation}\label{eq:rate_coeff}
  k_{j}(T) = A_{j}T^{b_{j}}\exp\pp{ -{\theta_{a,j}}/{T} },\qquad j
  = 1,\dots,M,
\end{equation}
where $A_{j}$ is a constant (often called pre-exponential), $b_{j}$ is the temperature exponent, and $\theta_{a,j}$ is the activation temperature.
Some reactions can display concentration dependence in their rate of progress, which is modeled via more complex expressions for their rate coefficient~\cite{ref:Troe1974}.
Pyrometheus includes both these and additional potential pressure dependencies.

The equilibrium constant $K_{j}$ in \cref{eq:reaction_rates} is evaluated through equilibrium thermodynamics
\begin{equation}\label{eq:equil_constants}
  K_{j}(T) = \pp{ \frac{p_{0}}{\mathscr{R}T} }^{\sum_{i =
      0}^{\nu_{ij}}}\exp\pp{ -\sum_{i =
      1}^{N}\frac{\nu_{ij}\hat{g}_{i}^{0}(T)}{\mathscr{R}T} },\qquad j
  = 1,\dots,M,
\end{equation}
where $p_{0} = 1\,\mathrm{atm}$, $\mathscr{R}$ is the universal gas constant, and
\begin{equation}\label{eq:gibbs}
  \hat{g}_{k}^{0}(T) = \hat{h}_{k}(T) - T\,\hat{s}_{k}^{0}(T),\qquad k
  = 1,\dots,N
\end{equation}
is the species standard Gibbs functions (in $\mathrm{J/mol}$), with $\{\hat{h}_{k}\}_{k = 1}^{N}$ and $\{\hat{s}_{k}^{0}\}_{k = 1}^{N}$ the species enthalpies and standard entropies (in $\mathrm{J/mol}$ and $\mathrm{J/mol \cdot K}$). These are modeled using NASA polynomials~\cite{ref:McBride2002}
\begin{align}\label{eq:nasa_poly}
  &\frac{\hat{h}_{i}}{\mathscr{R}T} = \frac{\alpha_{h}}{T} + \sum_{m =
    1}^{5}\frac{\alpha_{m}}{m}T^{m-1},\\ &\frac{\hat{s}_{i}^{0}}{\mathscr{R}}
  = \alpha_{s} + \alpha_{h}\log\,T + \sum_{m =
    2}^{5}\frac{\alpha_{m}}{m-1}T^{m},
\end{align}
where $\alpha_{h}$, $\alpha_{s}$, and $\{ \alpha_{m} \}_{m = 1}^{5}$ are the fit coefficients.

The presented combustion thermochemistry formulation needs to be solved numerically, even for small mechanisms such as hydrogen--air combustion.
Corresponding numerical methods are standard but judiciously selected to address numerical stiffness that arises as a consequence of the broad range of chemical time scales involved; these are summarized in \cref{subsec:time_integ}.
As will be shown, these methods make use of chemical Jacobians, so corresponding implementations used in \cref{sec:results} make use of the proposed computational approach that enables automatic differentiation, presented in full in \cref{sec:design}.

\subsection{Extensions to Flow \& Transport}\label{subsec:transport}

For problems that involve flow, \cref{eq:massfrac_ode} generalizes to
\begin{equation}\label{eq:massfrac_pde}
  \pop{}{t}\pp{\rho Y_{k}} + \pop{}{x_{j}}\pp{ \rho Y_{k} u_{j} } +
  \pop{\varphi_{kj}}{x_{j}} = S_{k},\qquad k = 1,\,\dots,\, N,
\end{equation}
with appropriate boundary conditions.
In \cref{eq:massfrac_pde}, $\mathbf{u} = \{ u_{j} \}_{j = 1}^{3}$ is the flow velocity, and $\boldsymbol{\varphi}_{k} = \{ \varphi_{kj}\}_{j = 1}^{3}$ the mass diffusion flux of the $k^{\mathrm{th}}$ species.
This term must be modeled; there are multiple options, a typical choice being the mixture-averaged approximation
\begin{equation}\label{eq:mixavg_flux}
  \varphi_{kj} = \varphi_{kj}^{\ast} + \varphi_{kj}^{c}
\end{equation}
where
\begin{equation}
   \varphi_{kj}^{\ast} = -\rho
   \mathscr{D}_{(k),m}\frac{W_{(k)}}{W}\pop{X_{k}}{x_{j}}
\end{equation}
is the mixture-averaged approximation, with $X_{k} = W Y_{k} /
W_{(k)}$ the mole fraction of the $k^{\mathrm{th}}$ species, and
\begin{equation}
  \varphi_{kj}^{c} = -Y_{k}\sum_{n = 1}^{N}\varphi_{nj}^{\ast}
\end{equation}
is a correction to ensure mass conservation.
In \cref{eq:mixavg_flux},
\begin{equation}\label{eq:diff_mix_rule}
  \mathscr{D}_{k,m} = \frac{W - X_{(k)}W_{k}}{W}\pp{ \sum_{j \neq
      k}\frac{X_{j}}{\mathscr{D}_{kj}} }^{-1}
\end{equation}
is the mixture-averaged diffusivity of species $k$, where the binary diffusivities $\mathscr{D}_{kj}$ are often approximated using polynomials:
\begin{equation}
  \mathscr{D}_{ij} = \sqrt{T}\sum_{m = 0}^{4}a_{i,j,m}\pp{\log T}^{m}.
\end{equation}

In \cref{eq:massfrac_pde}, the flow velocity $\mathbf{u}$ is obtained from the full set of conservation equations for mass, momentum, and energy density, provided elsewhere~\cite{ref:Cisneros2022}.
These involve two additional transport properties that must be modeled: the mixture viscosity $\mu$ and thermal conductivity $\kappa$.
In the mixture-averaged formulation, the viscosity is
\begin{equation}\label{eq:mixavg_viscosity}
  \mu = \sum_{k = 1}^{N}\frac{X_{k}\mu_{k}}{\sum_{j =
      1}^{N}\Phi_{kj}X_{j}},
\end{equation}
where
\begin{equation}
  \Phi_{kj} = \frac{ \psq{ 1 + \sqrt{ \pp{
          \frac{\mu_{k}}{\mu_{j}}\sqrt{W_{j}/W_{k}} } } }^{2}
  }{\sqrt{8\pp{ 1 + W_{k}/W_{j}}}},
\end{equation}
and
\begin{equation}
  \mu_{i} = \sqrt{T}\psq{\sum_{m = 0}^{4}b_{i,m}\pp{\log T}^{m}}^{2}
\end{equation}
approximates the viscosity of species $k$.
The coefficients $b_{i,m}$ are obtained via least-squares fitting to collision integrals~\cite{ref:Kee2003}.
The thermal conductivity is
\begin{equation}\label{eq:mixavg_conductivity}
  \kappa = \frac{1}{2}\psq{\sum_{k = 1}^{N}X_{k}\kappa_{k} +
    \pp{\sum_{k = 1}^{N}\frac{X_{k}}{\kappa_{k}}}^{-1} },
\end{equation}
where the conductivity of individual species is also approximated using polynomials
\begin{equation}\label{eq:species_cond}
  \kappa_{i} = \sqrt{T}\sum_{m = 0}^{4}c_{i,m}\pp{\log T}^{m}.
\end{equation}
Material coefficients $\mathscr{D}_{k,m}$ \cref{eq:diff_mix_rule}, $\mu$ \cref{eq:mixavg_viscosity}, and $\kappa$ \cref{eq:mixavg_conductivity} are provided by the computational design of \cref{sec:design}, extending our computational capabilities to flow simulations.

\subsection{Numerical Methods for Combustion Thermochemistry}\label{sec:computation}

We next introduce numerical methods to solve the formulation in the preceding section.
Throughout their presentation, aspects that benefit from the proposed computational design of \cref{sec:design} are highlighted.
This section shows how to compute the temperature by inverting the internal energy via Newton iteration.
We then discuss time integration and discretization schemes for thermochemistry systems \cref{eq:massfrac_ode} and \cref{eq:massfrac_pde}.

\subsubsection{Temperature Inversion}\label{subsec:temp_newton}

Following \cref{subsec:gov_eqns}, the net production rates \cref{eq:production_rates} are expressed in terms of density $\rho$, temperature $T$, and mass fractions $\vec{Y} \equiv \{ Y_{i} \}_{i = 1}^{N}$.
Any $N + 2$ variables, not just $\{ \rho,\, T,\, \vec{Y} \}$, can define the state.
If the temperature is not explicitly solved for, such as in simulations of compressible flows, transported variables are the density $\rho$, the internal energy $e$, and the mass fractions $\vec{Y}$, so $T$ must be inverted from $\{ \rho,\, e,\, \vec{Y} \}$.
The internal energy per unit mass (in $\mathrm{J/kg}$) of an ideal gas mixture is
\begin{equation}\label{eq:int_energy}
  e = \vec{Y} \cdot \vec{e}(T),
\end{equation}
where $\vec{e} = \{ e_{k}(T) \}_{k = 1}^{N}$ are the species internal
energies
\begin{equation}
  e_{k}(T) = \frac{\hat{h}_{k}(T) - \mathscr{R}T}{W_{k}},\qquad k =
  1,\dots,N,
\end{equation}
with $\hat{h}_{k}$ species molar enthalpy (in
$\mathrm{J/kmol}$).
Thus, given $e$, \cref{eq:int_energy} provides an implicit equation for the temperature $T$.
In Pyrometheus, \cref{eq:int_energy} is inverted using a Newton method
\begin{equation}
  T_{m + 1} = T_{m} + \Delta T_{m}
\end{equation}
where the subscript indicates the Newton iteration, and
\begin{equation}\label{eq:delta_temp}
  \Delta T_{m} =
  -\psq{\vec{Y}\cdot\pop{\vec{e}}{T}\pp{T_{m}}}^{-1}\psq{
    \vec{Y}\cdot\vec{e}(T_{m}) -e } = -\frac{1}{c_{v}(T_{m},\,
      \vec{Y})}\psq{\vec{Y}\cdot\vec{e}(T_{m}) - e},
\end{equation}
with
\begin{equation}\label{eq:cv}
  c_{v} \equiv \frac{\partial e}{\partial T} = c_{p} - \mathscr{R}
\end{equation}
the mixture specific heat capacity at constant volume.
In \cref{eq:cv},
\begin{equation}
  c_{p} = \sum_{i = 1}^{N} c_{p,i}(T)Y_{i}
\end{equation}
is the mixture specific heat capacity at constant pressure; $c_{p,i}$, like the enthalpy and entropy of each species, is modeled with NASA polynomials
\begin{equation}\label{eq:nasa_poly_c}
  \frac{c_{p,i}}{\mathscr{R}} = \sum_{m =
    1}^{4}m\,\frac{\alpha_{m}}{m}T^{m-1}.
\end{equation}

In some cases, such as those involving characteristic boundary conditions or simulation of low-Mach number combustion, it is advantageous to use enthalpy $h$ rather than internal energy.
For these cases, the temperature can be analogously inverted from $h$ and $\vec{Y}$ with minor modifications to the preceding Newton procedure.

\subsubsection{Time Integration \& Spatial Discretization}\label{subsec:time_integ}

Pyrometheus is a generalized thermochemistry toolkit that can be used in combustion simulations without restrictions on the time integration method.
For the numerical experiments of \Cref{sec:results}, specific time integration schemes are implemented.
The following discusses these and demonstrates in more detail how to develop a solver around Pyrometheus if necessary.

The stiff ODE system \cref{eq:massfrac_ode} is discretized via the Crank--Nicolson method
\begin{equation}\label{eq:crank_nicol_rhs}
  \underbrace{ Y_{k}^{m+ 1} - Y_{k}^{m} - \frac{\Delta t}{2}\pp{
      S_{k}^{m + 1} + S_{k}^{m} } }_{\equiv \mathscr{F}_{k}} = 0,
\end{equation}
where the superscript indicates the time step.
The scheme \cref{eq:crank_nicol_rhs} is nonlinear through $S_{k}^{m + 1}$, so $Y_{k}^{m + 1}$ is obtained via Newton iterations
\begin{equation}
  Y_{k,\, n + 1}^{m + 1} = Y_{k,\, n}^{m + 1} + \Delta Y_{k,\, n}^{m +
    1},
    \label{eq:crank_nicol_newton_system}
\end{equation}
where the second subscript indicates the $n^{\mathrm{th}}$ Newton iteration and $\Delta Y_{k,\, n}^{m + 1}$ are the Newton updates.
These are the solutions to
\begin{equation}\label{eq:crank_nicol_newton_update}
  \mathscr{J}_{jk,\, n}^{m + 1}\, \Delta{Y}_{k,\, n}^{m + 1} =
  -\mathscr{F}_{k,\, n}.
\end{equation}
where
\begin{equation}\label{eq:crank_nicol_jac}
  \mathscr{J}_{jk,\, n}^{m + 1} \equiv \frac{\partial
    \mathscr{F}_{j,\, n}}{\partial Y_{k,\, n}^{m + 1}} = \delta_{jk} -
  \frac{\Delta t}{2}J_{jk,\, n}^{m + 1},
\end{equation}
with
\begin{equation}\label{eq:chem_jacobian}
    J_{jk} \equiv \frac{\partial S_{j}}{Y_{k}},\qquad j = 1,\,\dots,\,\ N,\,\text{ and }\, k = 1,\,\dots,\,N
\end{equation}
the chemical Jacobian (evaluated at $\{Y_{k,\, n}\}_{k = 1}^{N}$), and $\delta_{jk}$ the Kronecker delta.
The Newton method for each time step is initialized with $Y_{k,\, 0}^{m + 1} = Y_{k}^{m}$, $k = 1,\,\dots,\,N$.

Simulations of reactors with flow, such as flame, entail spatial discretization of \cref{eq:massfrac_pde}.
As with time marching, Pyrometheus is independent of the choice of discretization scheme.
In subsequent demonstrations of \cref{subsub:app_sims}, we approximate spatial derivatives in \cref{eq:massfrac_pde} using standard fourth-order central stencils in the interior and one-sided second-order stencils near the boundaries.
Additionally, to suppress high wavenumber oscillations, an eight-order filter is applied after every time step. 

\bibliographystyle{bibsty}
\bibliography{ref.bib}

\end{document}